\newcommand{\printstyle}{reprint}
\newcommand{\halfwidth}{\columnwidth} 
\newcommand{\thirdwidth}{0.9\halfwidth} 
\newcommand{\fullwidth}{\textwidth}

\documentclass[aip,pop,floatfix,\printstyle ,nofootinbib]{revtex4-1}

\usepackage{color}

\usepackage[caption=false]{subfig}
\usepackage{graphicx,booktabs,array}
\usepackage{multirow}
\DeclareGraphicsExtensions{.pdf,.png,.jpg,.eps}
\usepackage{epsfig}
\usepackage{epstopdf}
\usepackage{amssymb}
\usepackage{amsmath}
\usepackage{amsfonts}
\usepackage{mathrsfs}
\usepackage{amsthm}
\usepackage{float}
\usepackage{amsbsy}
\usepackage{bm}
\usepackage{grffile}
\usepackage{hyperref}
\hypersetup{
    colorlinks = true,
    citecolor = blue,
    urlcolor = blue,
    linkcolor = blue,
}
\usepackage{courier}
\usepackage{upgreek}
\usepackage{xspace}
\usepackage{xcolor}
\usepackage{calc} 
\usepackage{soul} 

\newcommand{\ie}{\textit{i}.\textit{e}. }
\newcommand{\eg}{\textit{e}.\textit{g}. }

\newcommand{\Alfven}{Alfv\'{e}n\xspace}
\newcommand{\Alfvenic}{Alfv\'{e}nic\xspace}

\newcommand{\va}{v_A}
\newcommand{\vs}{v_S}
\newcommand{\vb}{v_0}
\newcommand{\vc}{v_c}

\newcommand{\omegaci}{\omega_{ci}}

\newcommand{\omegacio}{\omega_{ci0}} 
\newcommand{\omegator}{\omega_\phi}
\newcommand{\omegapol}{\omega_\theta}
\newcommand{\omeganorm}{\omega / \omegaci}

\newcommand{\omegabar}{\bar{\omega}}
\newcommand{\omegace}{\omega_{ce}}
\newcommand{\omegape}{\omega_{pe}}

\newcommand{\ldebye}{\lambda_D}

\newcommand{\db}{\delta B}

\newcommand{\dbpar}{\db_\parallel}

\newcommand{\dbperp}{\db_\perp}
\newcommand{\Epar}{E_\parallel}
\newcommand{\Eperp}{E_\perp}

\newcommand{\kpar}{k_\parallel}
\newcommand{\kperp}{k_\perp}

\newcommand{\pphi}{p_\phi}
\newcommand{\pmin}{p_{\text{min}}}

\newcommand{\vinj}{\vb/\va}

\newcommand{\linj}{\lambda_0}
\newcommand{\dl}{\Delta\lambda}
\newcommand{\dlinv}{\dl^{-1}}

\newcommand{\dfhym}{\delta f}
\newcommand{\vcrit}{\vc/\vb}

\newcommand{\nbo}{n_b}
\newcommand{\neo}{n_e}
\newcommand{\nb}{\nbo / \neo}

\newcommand{\vpar}{v_\parallel}
\newcommand{\vpres}{v_{\parallel,\text{res}}}

\newcommand{\vdrift}{v_{\text{Dr}}}
\newcommand{\vperp}{v_\perp}

\newcommand{\lres}{\ell}

\newcommand{\betae}{\beta_e}
\newcommand{\nperp}{n_\perp}
\newcommand{\npar}{n_\parallel}
\newcommand{\Nperp}{N_\perp}
\newcommand{\Npar}{N_\parallel}
\newcommand{\vtherme}{v_{th,e}}

\newcommand{\W}{\mathcal{E}}

\newcommand{\vE}{\vec{E}}
\newcommand{\vB}{\vec{B}}

\newcommand{\vk}{\vec{k}}

\newcommand{\vJ}{\vec{J}}
\newcommand{\vV}{\vec{V}}

\newcommand{\xinj}{x_0}
\newcommand{\dx}{\Delta x}

\newcommand{\fb}{f_0}
\newcommand{\fl}{f_L} 
\newcommand{\fbeam}{f_b}
\newcommand{\omegacires}{\avg{\bar{\omega}_{ci}}}

\newcommand{\kratraw}{\kpar/\kperp}
\newcommand{\krat}{\abs{\kratraw}}
\newcommand{\kratinv}{\abs{\kperp/\kpar}}
\newcommand{\rhob}{\rho_{\perp b}}

\newcommand{\zp}{\zeta}

\newcommand{\eqlab}[1]{\quad\text{#1}}

\newcommand{\Te}{T_e}

\newcommand{\lm}{\lambda_m}

\newcommand{\dv}{\Delta v}

\newcommand{\Ksym}{K}
\newcommand{\Kij}{\Ksym_{ij}}
\newcommand{\Kxx}{\Ksym_{11}}
\newcommand{\Kxy}{\Ksym_{12}}

\newcommand{\Kyx}{\Ksym_{21}}
\newcommand{\Kyy}{\Ksym_{22}}
\newcommand{\Kyz}{\Ksym_{23}}

\newcommand{\Kzy}{\Ksym_{32}}
\newcommand{\Kzz}{\Ksym_{33}}
\newcommand{\antiherm}{^{A}}
\newcommand{\herm}{^{H}}

\newcommand{\Wwave}{\mathcal{W}}
\newcommand{\Pwave}{\mathcal{P}}
\newcommand{\gammadamp}{\gamma_{\text{damp}}}

\newcommand{\alphacrithym}{\krat_\text{crit}}

\newcommand{\pphipow}{\sigma}

\newcommand{\avg}[1]{\left\langle #1 \right\rangle}
\renewcommand{\vec}[1]{\bm{#1}}
\newcommand{\ten}[1]{\cdot 10^{#1}}
\newcommand{\abs}[1]{\left|#1\right|}
\renewcommand{\dot}{\cdot}
\newcommand{\cross}{\times}
\renewcommand{\div}{\nabla\dot}
\newcommand{\grad}{\nabla}
\newcommand{\curl}{\nabla\cross}
\newcommand{\defined}{\equiv}
\newcommand{\like}{\sim}
\newcommand{\conj}{{}^*}
\newcommand{\ord}[1]{\mathcal{O}\left(#1\right)}

\newcommand{\approptoinn}[2]{\mathrel{\vcenter{
  \offinterlineskip\halign{\hfil$##$\cr
    #1\propto\cr\noalign{\kern2pt}#1\sim\cr\noalign{\kern-2pt}}}}}
\newcommand{\appropto}{\mathpalette\approptoinn\relax}

\newcommand{\tto}{\text{ to }}

\renewcommand{\Re}[1]{\text{Re}\left[#1\right]}
\renewcommand{\Im}[1]{\text{Im}\left[#1\right]}

\newcommand{\pderiv}[2]{\frac{\partial #1}{\partial #2}}

\newcommand{\figref}[1]{Fig.\xspace\ref{#1}}
\renewcommand{\eqref}[1]{Eq.\xspace\ref{#1}}
\newcommand{\secref}[1]{Sec.\xspace\ref{#1}}
\newcommand{\citeref}[1]{Ref.\xspace\onlinecite{#1}}
\newcommand{\appref}[1]{Appendix\xspace\ref{#1}}

\newcommand{\code}[1]{\texttt{#1}\xspace}
\newcommand{\HYM}{\code{HYM}}
\newcommand{\TRANSP}{\code{TRANSP}}
\newcommand{\NUBEAM}{\code{NUBEAM}}

\newcommand{\CAETB}{\code{CAE3B}}

\newcommand{\myname}{J.B. Lestz} 
\newcommand{\Elena}{E.V. Belova} 
\newcommand{\Nikolai}{N.N. Gorelenkov}

\newcommand{\PPPL}{Princeton Plasma Physics Lab, Princeton, NJ 08543, USA}

\newcommand{\Princeton}{Department of Astrophysical Sciences, Princeton University, Princeton, NJ 08543, USA}

\newcommand{\UCIrvine}{Department of Physics and Astronomy, University of California, Irvine, CA 92697, USA}
\newcommand{\UCI}{\UCIrvine}

\newcommand{\Podesta}{Podest{\`a}}
\newcommand{\Fulop}{F{\"u}l{\"o}p}

\definecolor{darkgreen}{rgb}{0,0.5,0}

\begin{document}

\title{Hybrid simulations of sub-cyclotron compressional and global \Alfven Eigenmode stability in spherical tokamaks}

\author{\myname}
\email{jlestz@uci.edu}
\affiliation{\UCI}
\affiliation{\Princeton}
\affiliation{\PPPL}
\author{\Elena} 
\affiliation{\PPPL}
\author{\Nikolai}
\affiliation{\PPPL}
\date{\today}
\begin{abstract}

A comprehensive numerical study has been conducted in order to investigate the stability of beam-driven, sub-cyclotron frequency compressional (CAE) and global (GAE) \Alfven Eigenmodes in low aspect ratio plasmas for a wide range of beam parameters. The presence of CAEs and GAEs has previously been linked to anomalous electron temperature profile flattening at high beam power in NSTX experiments, prompting further examination of the conditions for their excitation. Linear simulations are performed with the hybrid MHD-kinetic initial value code \HYM in order to capture the general Doppler-shifted cyclotron resonance that drives the modes. Three distinct types of modes are found in simulations -- co-CAEs, cntr-GAEs, and co-GAEs -- with differing spectral and stability properties. The simulations reveal that unstable GAEs are more ubiquitous than unstable CAEs, consistent with experimental observations, as they are excited at lower beam energies and generally have larger growth rates. Local analytic theory is used to explain key features of the simulation results, including the preferential excitation of different modes based on beam injection geometry and the growth rate dependence on the beam injection velocity, critical velocity, and degree of velocity space anisotropy. The background damping rate is inferred from simulations and estimated analytically for relevant sources not present in the simulation model, indicating that co-CAEs are closer to marginal stability than modes driven by the cyclotron resonances. 

\end{abstract}
\maketitle

\section{Introduction}
\label{sec:Intro}

High frequency $(\omega \lesssim \omegaci)$ compressional (CAE) and global (GAE) \Alfven Eigenmodes are routinely driven unstable by super-\Alfvenic beam ions in the spherical tokamaks NSTX 
 \cite{Fredrickson2001PRL,Gorelenkov2003NF,Fredrickson2004POP,Crocker2011PPCF,Fredrickson2012NF,Fredrickson2013POP,Crocker2013NF,Crocker2018NF},
 NSTX-U \cite{Fredrickson2018NF,Kaye2019NF}, and MAST \cite{Appel2008PPCF,Sharapov2014POP}. They have also been observed in the conventional aspect ratio tokamaks DIII-D \cite{Heidbrink2006NF,Tang2021U} and ASDEX Upgrade \cite{Ochoukov2020NF}, with marginally \Alfvenic or sub-\Alfvenic fast ions. The CAEs/GAEs have been observed in the broad frequency range $0.1 < \omeganorm < 1.2$ with a wide range of toroidal mode numbers $\abs{n} = 1 - 15$, and may propagate toroidally either with (co-propagation) or against (cntr-propagation) the direction of the plasma current and beam injection. 
  
  The CAE, also known as the fast magnetosonic wave, has the simple dispersion relation $\omega_\text{CAE} \approx k \va$ in a uniform slab in the limit of zero plasma pressure, where $\va$ is the \Alfven speed ($\va^2 = B_0^2 / \mu_0 \rho_0$) and $B_0,\rho_0$ are the background magnetic field and plasma mass density, respectively. Its polarization is dominantly compressional with $\dbpar \gg \dbperp$. In toroidal geometry, the presence of a magnetic well on the low field side generates an effective potential well, yielding discrete eigenmodes trapped within the well \cite{Kolesnichenko1998NF,Gorelenkova1998POP,Gorelenkov2002POP,Smith2003POP,Gorelenkov2006NF}.
  
  The GAE is a specific type of shear \Alfven wave existing in non-uniform plasmas, radially localized near an extremum in the \Alfven continuum (typically near regions of low magnetic shear) \cite{Appert1982PP,Mahajan1983PF}. Hence, its dispersion is roughly determined by $\omega_\text{GAE} \leq \left[\abs{\kpar} \va(r)\right]_{\text{min}}$ (with the magnetic field calculated on-axis). As shear waves, GAEs have dominantly shear polarization, with $\dbperp \gg \dbpar$. In low aspect ratio equilibria, the shear and compressional branches are more strongly coupled, especially in the edge, making it difficult to distinguish the modes in experiments \cite{Crocker2013NF,Belova2017POP}.
  
   CAEs and GAEs are MHD waves which may be driven unstable by free energy available in a resonant sub-population of energetic particles. In order to interact with the wave, particles must satisfy the general Doppler-shifted cyclotron resonance condition. Three types of modes with distinct properties were unstable in the simulations: co-CAEs, cntr-GAEs, and co-GAEs. The characteristics of these three modes will be compared throughout this work. Each is driven primarily by a different form of the general Doppler-shifted cyclotron resonance: the co-CAEs are driven by the Landau resonance, the cntr-GAEs by the ordinary cyclotron resonance, and the co-GAEs by the anomalous cyclotron resonance. The nature of the ordinary and anomalous cyclotron resonances necessitates using a full orbit model for the fast ions which drive these modes. The details of the specific resonances play an important role in determining the stability properties of the modes driven by them. 
  
 The presence of strong CAE/GAE activity has been linked to anomalous electron temperature profile flattening in beam-heated NSTX plasmas \cite{Stutman2009PRL,Ren2017NF}. The precise cause of this flattening remains an important unanswered question, which could imperil the future development of low aspect ratio tokamaks since it limits core electron temperatures. While two mechanisms have been proposed theoretically, neither has been sufficient to reproduce the experimental observations. The first involves mode conversion from a core-localized CAE/GAE to a kinetic \Alfven wave (KAW) at the \Alfven resonance location $\omega = \omega_A(r_0) = \kpar \va(r_0)$, effectively channeling energy from the core to edge \cite{Kolesnichenko2010PRL,Belova2015PRL}. The second mechanism is orbit stochastization, where the presence of sufficiently many modes can enhance radial heat transport by two orders of magnitude through diffusive processes \cite{Gorelenkov2010NF}. Improved understanding of the preferential conditions for CAE/GAE excitation is an essential step towards predicting electron temperature profiles in discharges with substantial neutral beam heating, which can drive these modes.   

The purpose of this study is to leverage hybrid kinetic-MHD simulations of realistic NSTX conditions in order to investigate how the linear stability properties of CAEs and GAEs depend on key fast ion properties. The most heavily investigated parameters are the beam injection geometry and normalized beam injection velocity $\vinj$. A comprehensive parameter scan of these quantities is performed for a wide range of toroidal mode numbers. The inferred stability boundaries and growth rate trends are interpreted with recently derived expressions for the local fast ion drive \cite{Lestz2020p1,Lestz2020p2}, yielding very reasonable agreement. The effect of gradients in the fast ion distribution with respect to canonical toroidal angular momentum (dominated by radial dependence), self-consistently included in the simulations but absent in the local theory used throughout the paper, is considered and found to resolve some discrepancies. Two additional fast ion distribution parameters are also briefly studied: the normalized critical velocity $\vcrit$ and degree of velocity space anisotropy. The level of CAE/GAE damping on the background plasma is addressed by scanning the beam density within the simulations and also estimating the thermal electron damping (absent from the simulation model) analytically. Put together, the combination of a large set of realistic simulations and a simplified analytic theory form a clear picture of the preferential conditions for CAE and GAE destabilization which can inform future experiments. 

The paper is organized as follows. The hybrid simulation model is described in \secref{sec:model}. In \secref{sec:id}, an overview of the basic properties of the different types of unstable modes in the simulations is given. \secref{sec:stabres} presents the results of a comprehensive parameter scan of the beam injection geometry and velocity, and interprets the growth rate dependence and spectral trends with a local analytic theory. In \secref{sec:vcdl}, a brief examination of the dependence of the growth rate on the normalized critical velocity and degree of velocity space anisotropy is presented. The relevance of plasma non-uniformity is considered in \secref{sec:pphi}. Lastly, \secref{sec:damp} evaluates the level of CAE/GAE damping on the background plasma. A discussion of the main results is given in \secref{sec:summary}. 

\section{Simulation Scheme}
\label{sec:model}

The simulations in this work are conducted using the 3D hybrid MHD-kinetic initial value code \HYM \cite{Belova1997JCP,Belova2017POP}. In this code, the thermal electrons and ions are modeled as a single resistive, viscous MHD fluid. The minority energetic beam ions $(\nbo \ll \neo)$ are treated kinetically with a full-orbit $\delta f$ scheme. When studying high frequency modes with $\omega \lesssim \omegaci$, resolving the fast ion gyromotion is crucial to capturing the general Doppler-shifted cyclotron resonance that drives the modes. The two species are coupled together using current coupling in the momentum equation below

\begin{align}
\rho \frac{d \vV}{d t} = -\nabla P + (\vJ - \vJ_b) \cross \vB - e n_b (\vE - \eta \delta\vJ) + \mu_v \Delta \vV
\label{eq:momentum}
\end{align}

Here, $\rho, \vec{V}, P$ are the thermal plasma mass density, fluid velocity, and pressure, respectively. The magnetic field can be decomposed into an equilibrium and perturbed part $\vB = \vB_0 + \delta\vB$, while the electric field $\vE$ has no equilibrium component. The beam density and current are $n_b$ and $\vec{J_b}$. The total plasma current is determined by $\mu_0\vec{J} = \curl \vec{B}$ while $\mu_0\delta\vec{J} = \curl \delta\vec{B}$ is the perturbed current. Non-ideal MHD physics are introduced through the viscosity coefficient $\mu_v$ and resistivity $\eta$. \eqref{eq:momentum} results from summing over thermal ion and electron momentum equations, neglecting the electron mass, and enforcing quasineutrality $n_e = n_b + n_i$. In addition to \eqref{eq:momentum}, the thermal plasma evolves according to the following set of fluid-Maxwell equations 

\begin{subequations}
\begin{align}
\label{eq:fluids1}
\vec{E} &= - \vec{V} \cross \vec{B} + \eta \delta\vec{J} \\
\label{eq:fluids2}
\frac{\partial \vec{B}}{\partial t} &= - \curl \vec{E} \\ 
\label{eq:fluids3}
\frac{\partial \rho}{\partial t} &= - \div \left( \rho \vec{V} \right) \\ 
\label{eq:fluids4}
\frac{d}{dt}&\left( \frac{P}{\rho^{\gamma_a}} \right) = 0 
\end{align}
\label{eq:fluids}
\end{subequations}

In fully nonlinear simulations, the pressure equation includes Ohmic and viscous heating in order to conserve the system's energy (see Eq. 2 of \citeref{Belova2017POP}). These effects are neglected in the linearized simulations presented here, reducing to the adiabatic equation of state in \eqref{eq:fluids4} with the adiabatic index $\gamma_a = 5/3$. Note that the terms $\vE - \eta\delta\vJ$ appear in \eqref{eq:momentum}, \eqref{eq:fluids1}, and \eqref{eq:parts2} due to quasineutrality and momentum conservation between the thermal plasma and beam ions. 

Field quantities are evolved on a cyclindrical grid, with particle quantities stored on a Cartesian grid sharing the $Z$ axis with the fluid grid. For simulations of $n < 8$, a field grid of $N_Z\times N_R\times N_\phi = 120 \times 120 \times 64$ is used. For larger $n$, the resolution is refined to $120 \times 96 \times 128$. The particle grid has $N_Z \times N_X \times N_Y = 120 \times 51 \times 51$, with at least $500,000$ particles used in each simulation. Convergence studies were conducted on the grid resolution and number of simulation particles, which can lead to slight variations in growth rate but no change in frequency or mode structure. 

The fast ion distribution is decomposed into an equilibrium and perturbed part, $\fbeam = \fb + \dfhym$. Each numerical particle has a weight $w = \dfhym / \fl$ where $\fl$ is a function of integrals of motion used for particle loading $(d\fl/dt = 0)$. The $\dfhym$ particles representing the fast ions evolve according to the following equations of motion

\begin{subequations}
\begin{align}
\label{eq:parts1}
\frac{d\vec{x}}{dt} &= \vec{v} \\
\label{eq:parts2}
\frac{d\vec{v}}{dt} &= \frac{q_i}{m_i}\left(\vec{E} - \eta \delta\vec{J} + \vec{v} \cross \vec{B}\right) \\ 
\label{eq:dwdt}
\frac{dw}{dt} &= -\left(\frac{\fbeam}{\fl} - w\right)\frac{d \ln \fb}{dt}
\end{align}
\label{eq:parts}
\end{subequations}

Particle weights are used to calculate the beam density $n_b$ and current $\vJ_b$ which appear in \eqref{eq:momentum}. The $\delta f$ scheme has two advantages: the reduction of numerical noise and intrinsic identification of resonant particles, which are those with largest weights at the end of the simulation. 

The equilibrium fast ion distribution function is written as a function of the constants of motion $\W$, $\lambda$, and $\pphi$. The first, $\W = \frac{1}{2}m_i v^2$, is the particle's kinetic energy. Next, $\lambda = \mu B_0 / \W$ is a trapping parameter, where first order corrections in $\rho_{EP}/L_B$ to the magnetic moment $\mu$ are kept for improved conservation in the simulations \cite{Belova2003POP}. This correction is more relevant in low aspect ratio devices since the fast ion Larmor radius can be a significant fraction of the minor radius, leading to nontrivial variation in $B_0$ during a gyro-orbit. To lowest order, one may re-write $\lambda \approx (\vperp^2/v^2)(\omegacio/\omegaci)$ such that in a tokamak, passing particles will have $0 < \lambda < 1 - r/R$ and trapped particles will have $1 - r/R < \lambda < 1 + r/R$. Hence, $\lambda$ is a complementary variable to a particle's pitch $\vpar/v$. Lastly, $\pphi = -q_i \psi + m_i R v_\phi$ is the canonical toroidal angular momentum, conserved due to the axisymmetric equilibria used in these simulations. Here, $\psi$ is the poloidal magnetic flux and $\psi_0$ is its on-axis value. A separable form of the beam distribution is assumed: \cite{Belova2003POP} $\fb (v,\lambda,\pphi) = C_f f_1 (v) f_2 (\lambda) f_3 (\pphi,v)$

\begin{subequations}
\begin{align}
\label{eq:F1}
f_1(v) &= \frac{1}{v^3 + v_c^3} \quad \text{ for } v < \vb \\
\label{eq:F2}
f_2(\lambda) &= \exp\left(-\left(\lambda - \linj\right)^2 / \dl^2\right) \\ 
\label{eq:F3}
f_3\left(\pphi,v\right) &= \left(\frac{\pphi - \pmin}{m_i R_0 v - q_i \psi_0 - \pmin}\right)^\pphipow \text{ for } \pphi > \pmin
\end{align}
\label{eq:F0}
\end{subequations}

The energy dependence, $f_1(v)$, is a slowing down function with injection velocity $\vb$ and critical velocity $\vc$ \cite{Gaffey1976JPP}. For $v > \vb$, $f_1(v) = \exp(-(v-\vb)^2/\dv^2)/(v^3 + \vc^3)$ is used to model a smooth, rapid decay near the injection energy with $\dv = 0.1 \vb$. A beam-like distribution in $\lambda$ is used for $f_2(\lambda)$, centered around $\linj$ with constant width $\dl$. In reality, $\dl$ increases at lower energies due to pitch angle scattering, as accounted for in \citeref{Belova2019POP}, but this effect was not included in this study. Characteristic profiles of beam density calculated by the global transport code \TRANSP \cite{Goldston1981JCP} and Monte Carlo fast ion module \NUBEAM \cite{Pankin2004CPC} motivate the ad-hoc form of $f_3(\pphi,v)$. A prompt-loss boundary condition at the last closed flux surface is imposed by requiring $\pphi > \pmin = -0.1 q_i \psi_0$. Lastly, $C_f$ is a normalization constant determined numerically to match the peak of $\nb$ to its desired value. Values for the parameters appearing in the distribution are set in order to match the distribution in \NUBEAM from the target discharge. In this case, this yields $\vc = \vb/2$, $\vinj = 4.9$, $\linj = 0.7$, $\dl = 0.3$, $\sigma = 6$, and $n_b/n_e = 5.3\%$ as the baseline parameters. 

Importantly, \HYM includes the energetic particles self-consistently when solving for the equilibrium. Although the beam density is small, $\nbo \ll \neo$, the current carried by the beam can nevertheless be comparable to the thermal plasma current due to the large fast ion energy. Because this study is concerned with linear stability, the beam ions are evolved along their unperturbed equilibrium trajectories. Nonlinear simulations of CAEs \cite{Belova2017POP} and GAEs \cite{Belova2019POP} have also been conducted with the \HYM code. Since the modes in the linear simulations performed here do not saturate, the most linearly unstable mode will dominate and obscure all other modes. In order to study all of the eigenmodes, each toroidal mode number $n$ is simulated separately. 

In this study, approximately 600 simulations are performed to scan over the normalized injection velocity $\vinj = 2 - 6$ and injection geometry $\linj = 0.1 - 0.9$ of the beam ion distribution. The operating range for $\vinj$ in NSTX is approximately the same as the scanned range, while $\linj$ is restricted to approximately $0.5-0.7$ in experiment. The new, off-axis NSTX-U beam sources \cite{Gerhardt2012NF} have much more tangential injection, with $\linj \approx 0$. 

The bulk plasma equilibrium properties in these simulations are based on the well-diagnosed NSTX H-mode discharge 141398, which had $n_e = 6.7\ten{19}$ m$^{-3}$ and $B_0 = 0.325$ T on axis, with 6 MW of 90 keV beams corresponding to $\vinj = 4.9$, centered on $\linj \approx 0.7$. The on-axis ion cyclotron frequency was $f_{ci} = 2.5$ MHz. This plasma was chosen as the nominal scenario to study due to its rich spectrum of well-diagnosed high frequency modes and substantial amount of existing prior experimental analysis \cite{Crocker2013NF,Fredrickson2013POP,Crocker2018NF}.

\begin{figure}[tb]
\includegraphics[width = \thirdwidth]{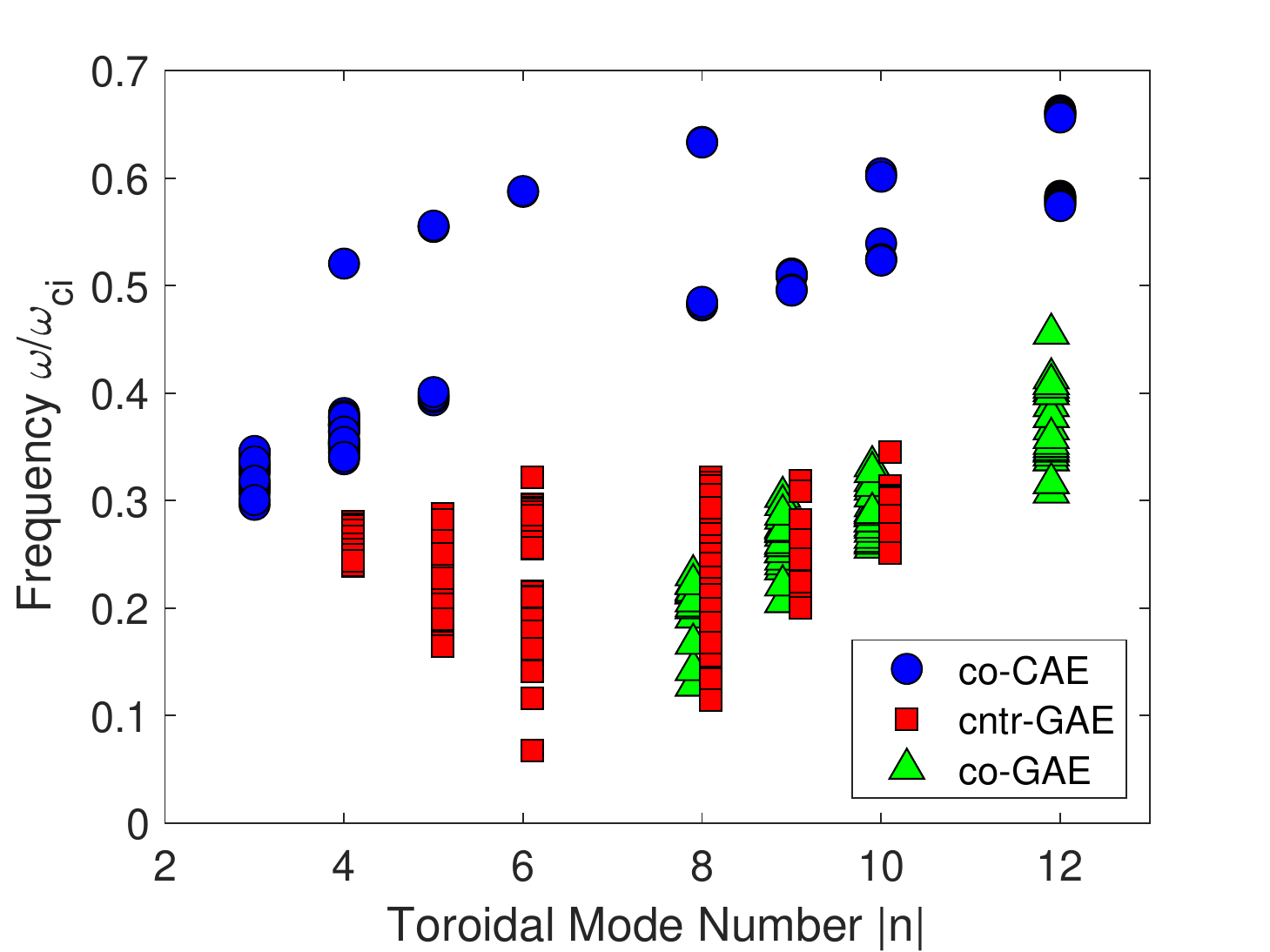}
\caption{Frequency of each type of mode as a function of toroidal mode number in simulations.}
\label{fig:frequency_vs_n}
\end{figure}

A perfectly conducting boundary condition is imposed at the boundary of the simulation volume. Since the shape of this boundary is not identical to the irregular shape of the NSTX(-U) vessel, the location of the boundary condition imposed in the simulations is different than it is in experiments. Previous simulations have shown that a smaller distance between the last closed flux surface and the bounding box can decrease the growth rate, so this discrepancy could lead to quantitative differences. However, this is not a concern for the goals of this study since it should not affect the trends in the growth rate with fast ion parameters. 

\section{Identification of Modes and Structures}
\label{sec:id} 

\begin{figure}[tb]
\includegraphics[width = \thirdwidth]{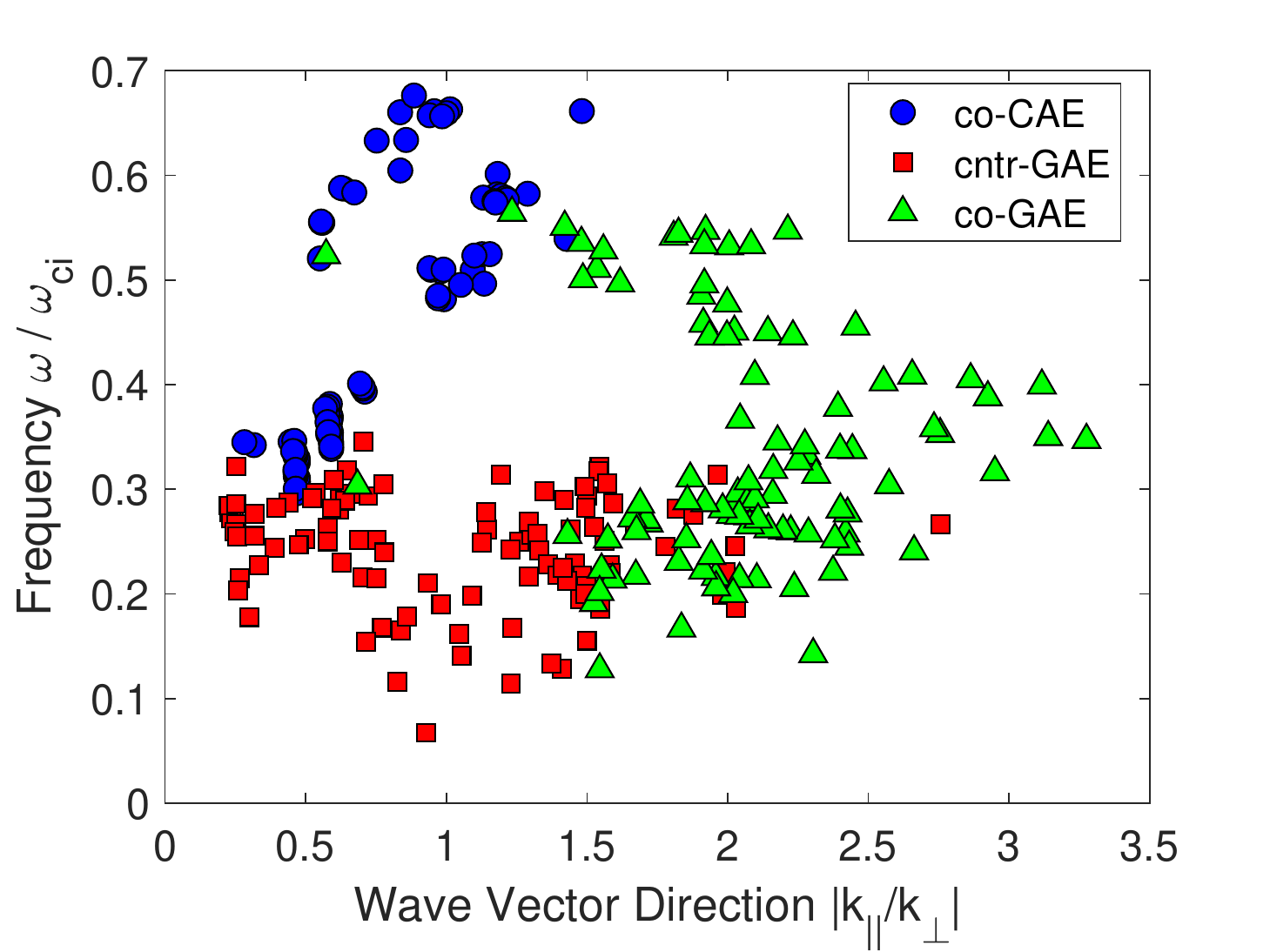}
\caption{Frequency and wave vector directions calculated from unstable modes in simulations for beam parameters $0.1 \leq \linj \leq 0.9$ and $2.5 \leq \vinj \leq 6$.}
\label{fig:ao_prop}
\end{figure}

Three distinct types of modes are found in this simulation study: co-propagating CAEs, counter-propagating GAEs, and co-propagating GAEs. \figref{fig:frequency_vs_n} summarizes the frequency range of each of these modes for the simulated toroidal mode numbers, while \figref{fig:ao_prop} shows the ranges of $\omeganorm$ and $\krat$ of each mode. Note that while $\kpar$ is often inferred from the large tokamak expression $\kpar = (n - m / q) / R$, this relation was not applied here due to the low aspect ratio of NSTX and ambiguous poloidal mode numbers in simulations. Instead, $\kpar$ was determined by taking a Fourier transform along equilibrium field lines traced on the flux surface with largest RMS fluctuation magnitude in $\dbpar$ for CAEs and a component of $\dbperp$ for GAEs. Meanwhile, peaks in the spatial Fourier transforms in $Z$, $R$, and $\phi$ give $k = \sqrt{k_R^2 + k_Z^2 + k_\phi^2}$, which can be used to infer $\kperp = \sqrt{k^2 - \kpar^2}$. This section will detail the characteristics of each simulated mode and contrast their properties. Distinguishing between CAEs and GAEs is notoriously difficult \cite{Crocker2013NF} in experiments due to similar frequency ranges and mixed polarization of the modes at the outboard side where magnetic coils are available for measurements. Hence, the wealth of information provided by simulations is valuable in guiding future experimental analysis. 

Fast ions can interact with the waves through the general Doppler shifted cyclotron resonance condition: 

\begin{align}
\omega - \avg{\kpar\vpar} - \avg{\kperp\vdrift} = \lres\avg{\omegaci}
\label{eq:reskpar}
\end{align} 

Here, $\avg{\dots}$ denotes orbit-averaging. The cyclotron resonance coefficient $\lres = -1,0,1$ for the frequency range studied here, though a resonance may exist for any integer value. The Landau resonance corresponds to $\lres = 0$, the ``ordinary'' cyclotron resonance has $\lres = 1$, and the ``anomalous'' cyclotron resonance has $\lres = -1$. Note that for sub-cyclotron frequencies, and in the usual case where $\abs{\avg{\kpar\vpar}} \gtrsim \abs{\avg{\kperp\vdrift}}$, counter-propagating modes $(\kpar < 0)$ can only satisfy the ordinary cyclotron resonance, while co-propagating modes can interact through the Landau or anomalous cyclotron resonances, depending on their frequency. \eqref{eq:reskpar} can be equivalently written in terms of orbital frequencies: 

\begin{align}
\omega - n\avg{\omegator} - p\avg{\omegapol} = \lres\avg{\omegaci}
\label{eq:resomega}
\end{align}

In this expression, $\omegator$ and $\omegapol$ are the toroidal and poloidal orbital frequencies. The integer $n$ is the toroidal mode number (positive for co-propagation, negative for cntr-propagation) and $p$ is an integer. During simulations, $\avg{\omegator}$, $\avg{\omegapol}$, and $\avg{\omegaci}$ are computed for each fast ion, which enables determination of $p$ for the resonant particles, and hence identification of the dominant resonance in each simulation. 

\subsection{Co-propagating CAEs} 
\label{sec:idcae}

For a chosen set of beam parameters, our simulations find co-propagating CAEs for $n \geq 3$ with $0.28\omegaci < \omega < 0.70\omegaci$, marked by blue circles on \figref{fig:frequency_vs_n}. For all toroidal mode numbers, the CAE frequencies are larger than those of any unstable GAEs. This is reasonable since CAEs have frequencies proportional to $k$ whereas the GAE frequencies scale with $\abs{\kpar} < k$. Moreover, \figref{fig:ao_prop} shows that the CAEs have $\krat = 0.3 - 1.5$, with larger $\krat$ corresponding to the modes with higher $n$ numbers. As will also be the case with the GAEs, it is important to note that $\krat$ can range from small to order unity, violating the $\kpar\ll\kperp$ assumption that is often made in previous theoretical work that had large aspect ratio tokamaks in mind. This observation motivated in part the reexamination of the instability conditions for CAEs and GAEs in \citeref{Lestz2020p1} and \citeref{Lestz2020p2}, which will be used to interpret these simulation results in \secref{sec:simres}. 

\begin{figure}[tb]
\includegraphics[width = \halfwidth]{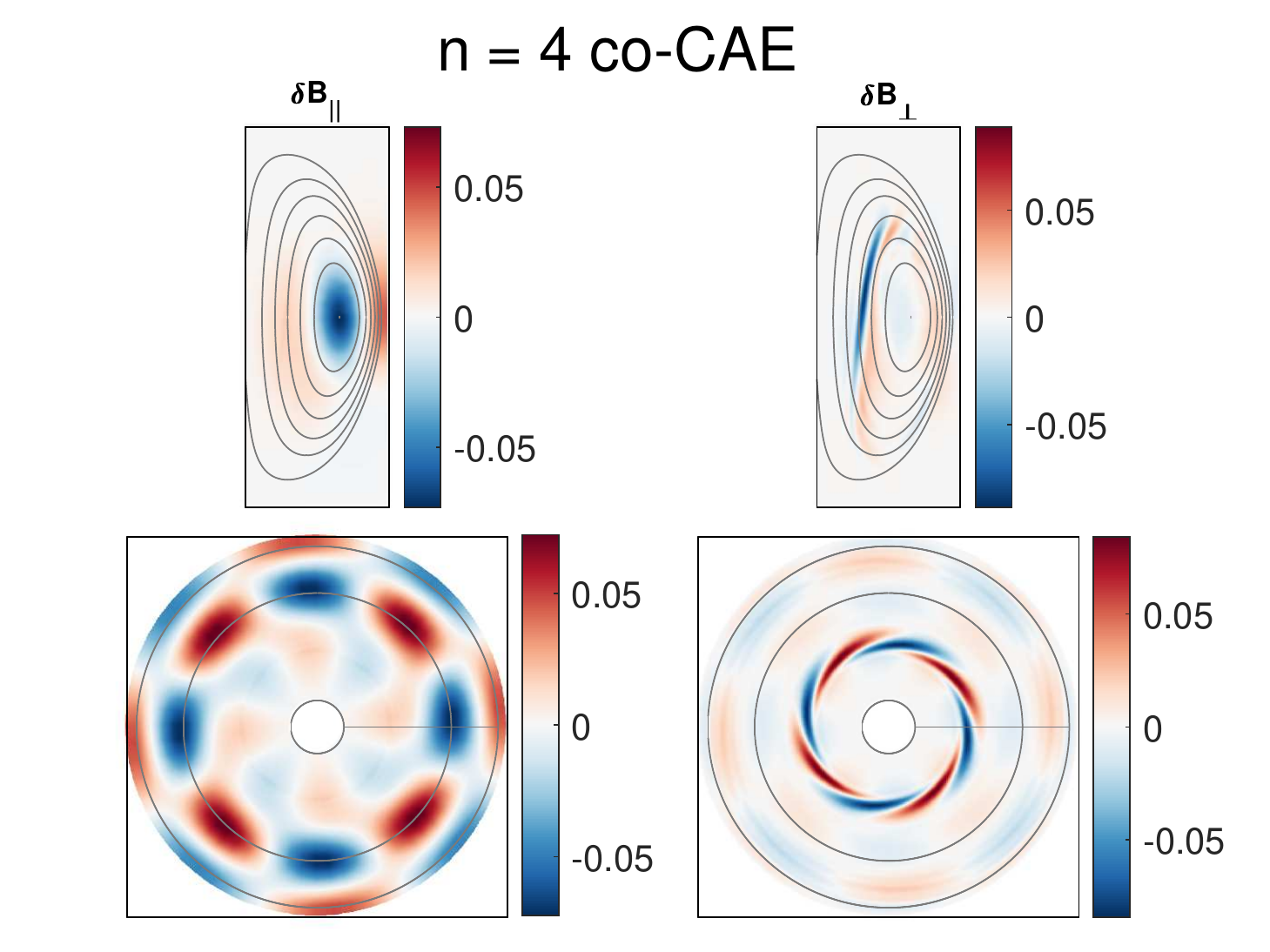}
\caption{Mode structure of an $n = 4$ co-CAE for $\vinj = 5.0$, $\linj = 0.7$. Top row is a poloidal cut, where gray lines are $\psi$ contours. Bottom row is a toroidal cut at the midplane, where gray lines represent the high field side, low field side, and magnetic axis. The second column is an arbitrary orthogonal component of $\dbperp$.}
\label{fig:CAEn4l0.7v5.0}
\end{figure}

This group of modes was identified as CAEs since they are high frequency sub-cyclotron modes with magnetic fluctuation dominated by the compressional component near the core, where $\dbpar/\dbperp \gg 1$. Qualitatively, $\dbpar$ for the low to moderate $n$ modes ($n < 10$) usually peaks on axis with low poloidal mode number $(m = 0 - 2)$. A typical example is shown in \figref{fig:CAEn4l0.7v5.0}, which shows an $n = 4$ co-propagating CAE with beam distribution parameterized by $\vinj = 5.0$ and $\linj = 0.7$ -- the parameters most closely matching the conditions of the NSTX discharge which this set of simulations are modeling. The core-localization of these modes agrees with previous nonlinear HYM simulations \cite{Belova2017POP}, contrasting with the analytic studies of CAEs under large aspect ratio assumptions, which predict localization near the edge \cite{Smith2003POP}. 

These modes also exhibit a substantial $\dbpar$ fluctuation on the low field side beyond the last closed flux surface, which is also a generic feature of counter-propagating GAE (see \figref{fig:GAEMn6l0.7v5.0}). This similarity has complicated previous efforts to delineate between high frequency AEs in experiment \cite{Crocker2013NF}. While $\dbperp$ is very small in the core for linear simulations dominated by a single CAE, there can still be large $\dbperp$ closer to the edge. This structure is most prominent on the high field side, as shown in \figref{fig:CAEn4l0.7v5.0}, but is also visible to a lesser degree on the low field side. The feature has been previously identified as a kinetic \Alfven wave (KAW) \cite{Belova2015PRL,Belova2017POP}, located at the \Alfven resonance location $\omega = \kpar \va(R,Z)$ where CAEs undergo mode conversion. The KAW appears whenever a CAE is unstable in these simulations. 

\begin{figure}[tb]
\includegraphics[width = \halfwidth*\real{0.8}]{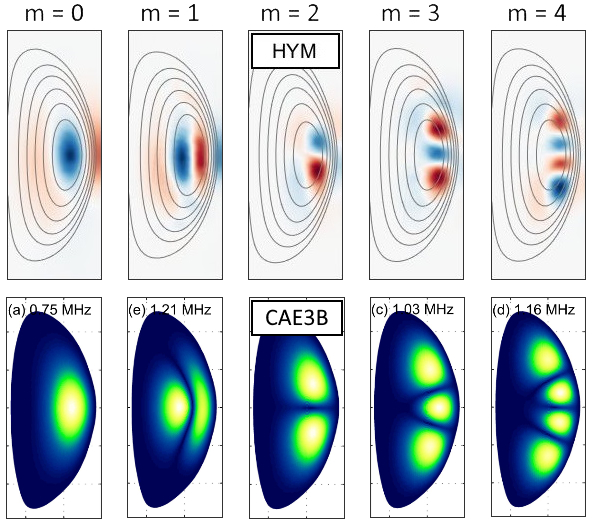} 
\caption{Top row: representative examples of poloidal mode structure ($\dbpar$) of co-CAEs from HYM simulations, labeled with qualitative poloidal mode numbers. From left to right, these modes have the following toroidal mode number and beam parameters: (1) $n=4, \linj = 0.7, \vinj = 5.0$, (2) $n = 5, \linj = 0.3, \vinj = 5.5$, (3) $n = 12, \linj = 0.9, \vinj = 5.1$, (4) $n =  10, \linj = 0.7, \vinj = 5.0$, (5) $n = 12, \linj = 0.7, \vinj = 5.1$. Bottom row: representative examples of poloidal mode structure of $n = -3$ cntr-CAEs from the spectral Hall MHD code \CAETB, solved with an equilibrium based on NSTX discharge 130335. The bottom row is adapted from Fig. 3 of \citeref{Smith2017PPCF}, reproduced with the permission of IOP publishing.}
\label{fig:CAE_HYM_struct}
\end{figure}

The modes with higher $n$ are localized closer to the edge on the low field side and have somewhat higher poloidal mode number ($m = 2 - 4$). The full range of poloidal mode structures of CAEs observed in linear HYM simulations is shown in \figref{fig:CAE_HYM_struct}. The first two modes, with $n = 4$ and $n = 6$, are more concentrated in the core, whereas the last three modes ($n = 12$, 10, and 12, respectively) are more localized near the edge. 

The labeling of these modes with poloidal mode numbers is somewhat arbitrary, as the modes do not lend themselves well to description with a single poloidal harmonic $m$ along the $\theta$ direction, as the structure can peak on axis or have structures that are poorly aligned with $\psi$ contours. This dilemma is also discussed in \citeref{Smith2009PPCF}, where the spectral code \CAETB is used to solve for CAEs at low aspect ratio within the Hall MHD model. Moreover, the CAE has a standing wave structure, indicating the presence of multiple signs of $m$, which in turn broadens the spectrum of $\kpar$ and $\kperp$ values for each mode. 

Regardless of the poloidal mode numbers ascribed to them, the CAE mode structures from the HYM simulations presented here qualitatively match the CAE eigenmodes found by the \CAETB eigensolver for a separate NSTX discharge 130335 \cite{Smith2017PPCF}. The similarity between the HYM and \CAETB mode structures provides further evidence that the instabilities seen in HYM and heuristically identified as CAE due to large $\dbpar$ in the core are indeed CAE solutions. Comparison against the CAE dispersion relation further supports the classification of these modes. In a uniform plasma, the magnetosonic dispersion including finite $\beta$ may be written 

\begin{equation}
\omega_{CAE}^2 = \frac{k^2 \va^2}{2}\left[1 + u^2 + \sqrt{\left(1 - u^2\right)^2 + \left(2u\kperp/k\right)^2}\right]
\label{eq:caebeta}
\end{equation}

Here, $u\defined \vs/\va = \sqrt{\gamma_a\beta/2}$ where $\gamma_a = 5/3$ is the adiabatic index and $\beta = 2\mu_0 P / B^2$. The finite $\beta$ corrections are important because the large fast ion pressure can make $\vs \like \va$ in the vicinity of the mode. In lieu of well-defined poloidal mode numbers, the mode structures from simulations are Fourier transformed to determined $k_R$, $k_Z$, and $k_\phi$. A quantitative agreement within $10\%$ is found between \eqref{eq:caebeta} and the mode frequencies in the simulations. 

\begin{figure}[tb]
\includegraphics[width = \halfwidth]{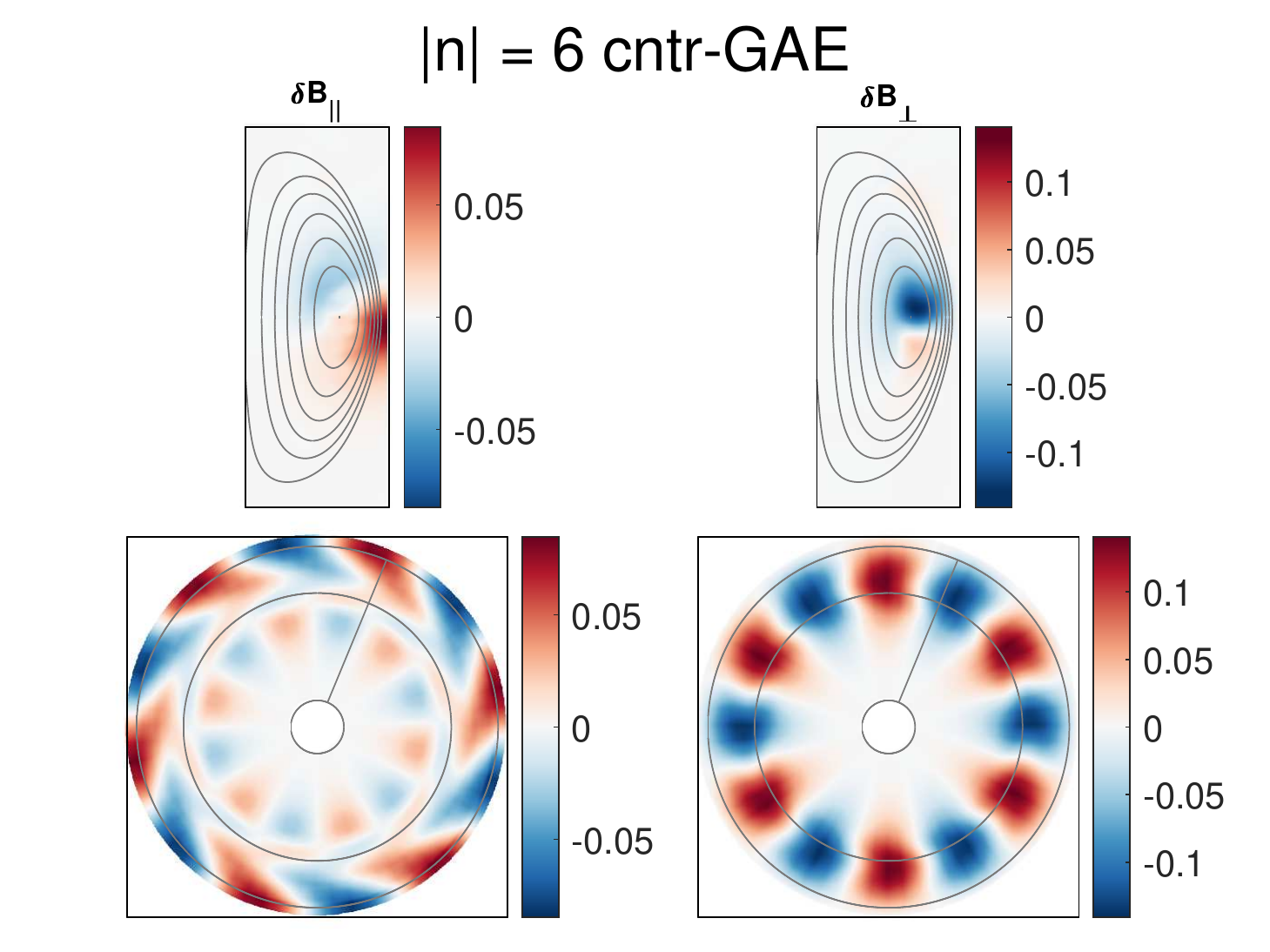}
\caption{Mode structure of an $n = -6$ for $\vinj = 5.0$, $\linj = 0.9$. Top row is a poloidal cut, where gray lines are $\psi$ contours. Bottom row is a toroidal cut at the midplane, where gray lines represent the high field side, low field side, and magnetic axis. The second column is an arbitrary orthogonal component of $\dbperp$.}
\label{fig:GAEMn6l0.7v5.0}
\end{figure}

The co-propagating CAEs are driven by the Landau resonance with fast ions: $\omega - n\avg{\omegator} - p\avg{\omegapol} = 0$. This condition has been verified in previous \HYM simulations \cite{Belova2017POP} by examining the fast ions with largest weights, which reveal the resonant particles due to \eqref{eq:dwdt}. Such examinations generally show that only a small number of resonances (\ie integers $p$) contribute nontrivially to each unstable mode -- a primary resonance $p$ and often two sub-dominant sidebands with $p\pm 1$. 

\begin{figure}[tb]
\includegraphics[width = \halfwidth*\real{0.8}]{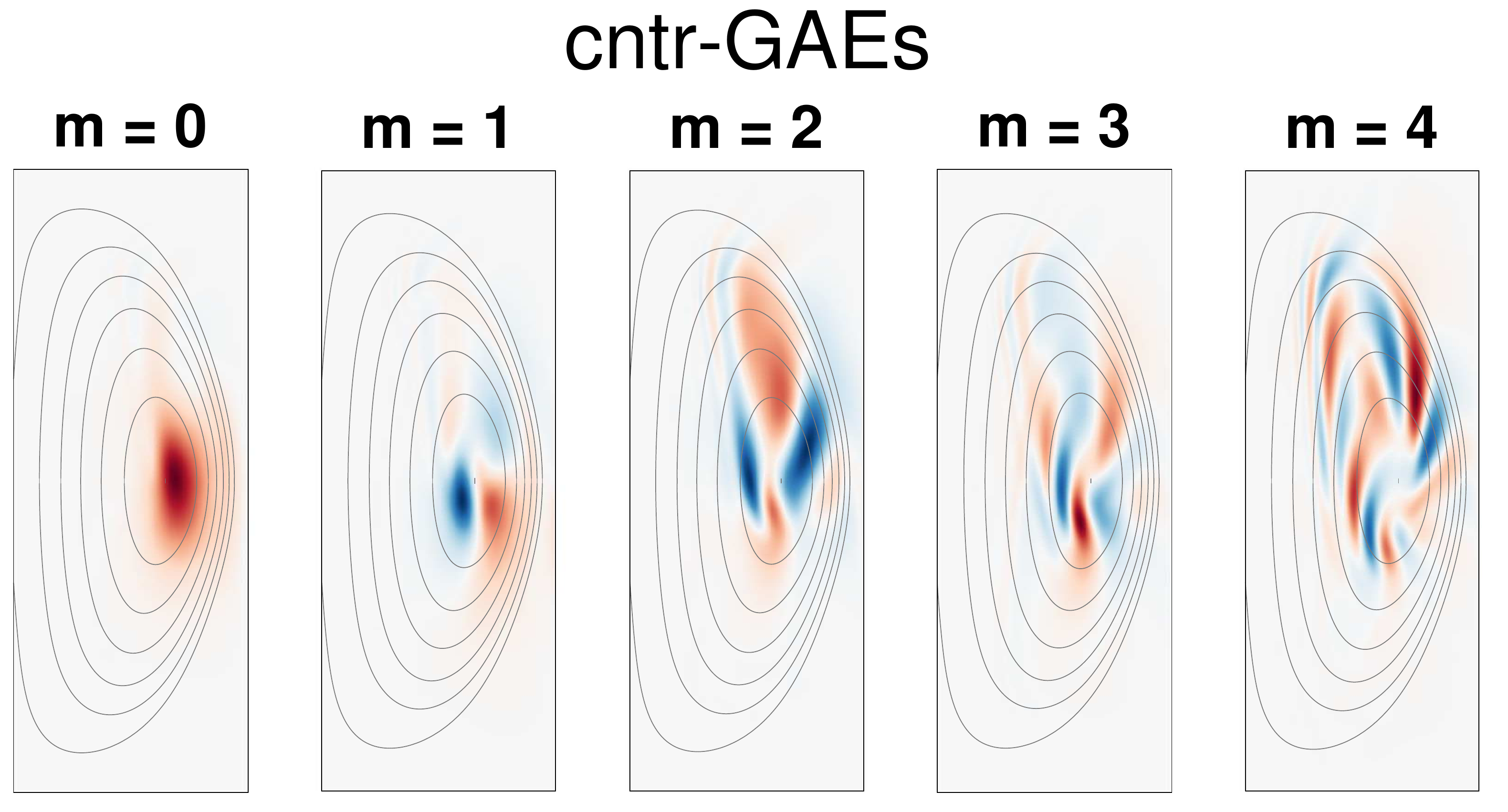}
\caption{Example poloidal mode structures $(\dbperp)$ of GAEs with different poloidal harmonics. From left to right, $m = 0$ ($n = 8, \linj = 0.9, \vinj = 4.5$), $m = 1$, ($n = 6, \linj = 0.7, \vinj = 4.5$), $m  = 2$ ($n = 6, \linj = 0.9, \vinj = 4.5$), $m = 3$ ($n = 5, \linj = 0.7, \vinj = 4.75$), and $m = 4$ ($n = 4, \linj = 0.9, \vinj = 5.0$).}
\label{fig:GAEMcomp}
\end{figure}

\subsection{Counter-propagating GAEs}
\label{sec:idgaem}

The global \Alfven eigenmodes in these simulations appeared in two flavors: co- and counter-propagating relative to the direction of the plasma current. The counter-propagating modes were excited for $\abs{n} = 4 - 10$ in the frequency range $0.05\omegaci < \omega < 0.35\omegaci$, and are indicated as red squares on \figref{fig:frequency_vs_n}. They exhibit a very wide range of wave vector directions, with unstable modes having $\krat = 0.2 - 3$ in the simulations.  

The cntr-GAEs were distinguished from CAEs due to their dominant shear polarization, \eg $\dbperp \gg \dbpar$ in the core where the fluctuation was largest, and their dispersion which generally scaled with the shear \Alfven dispersion $\omega \propto \kpar\va$. Counter-GAEs were routinely observed in NSTX \cite{Crocker2013NF} and NSTX-U \cite{Fredrickson2018NF} experiments with these basic characteristics, and previous comparisons between \HYM simulations and experimental measurements have revealed close agreement between the frequency of the most unstable counter-GAE for each $n$ \cite{Belova2019POP}. All counter-propagating modes which appeared in these simulations were determined to be GAEs, even though cntr-CAEs have also been observed in NSTX experiments \cite{Crocker2013NF}. This may be attributed to the initial value nature of these simulations, which are dominated by the most unstable mode. As discussed in \citeref{Lestz2020p1}, cntr-CAEs and cntr-GAEs are unstable for the same fast ion parameters, but the growth rate for the cntr-GAEs is typically larger. 

Similar to the CAEs, the counter-GAEs did not feature poloidal structure with well-defined mode numbers. Using an effective mode number $m$ loosely corresponding to the number of full wavelengths in the azimuthal direction, the GAEs ranged from $m = 0 - 5$, with lower $\abs{n}$ modes typically having larger $m = 3-5$ and modes with $\abs{n} > 7$ preferring smaller poloidal mode numbers of $m = 0 - 2$. Some representative examples are shown in \figref{fig:GAEMcomp}. The counter-GAE mode structure is typically more complex than that of the co-CAEs, likely due to the close proximity of the GAEs to the \Alfven continuum, which introduces shorter scale fluctuations on a kinetic scale that modulates the slower varying MHD structure. Comparing the mode structures in \figref{fig:CAE_HYM_struct} and \figref{fig:GAEMcomp}, one can see that while the CAEs are trapped in a potential well on the low field side \cite{Smith2003POP}, the GAEs can access all poloidal angles. The sheared mode structure present in \figref{fig:GAEMcomp} may be due to nonperturbative energetic particle effects, which have previously been documented in experimental observations \cite{Tobias2011PRL} and simulations \cite{Spong2013NF,Wang2015POP} of shear \Alfven waves in tokamaks. 

\begin{figure}[tb]
\includegraphics[width = \halfwidth]{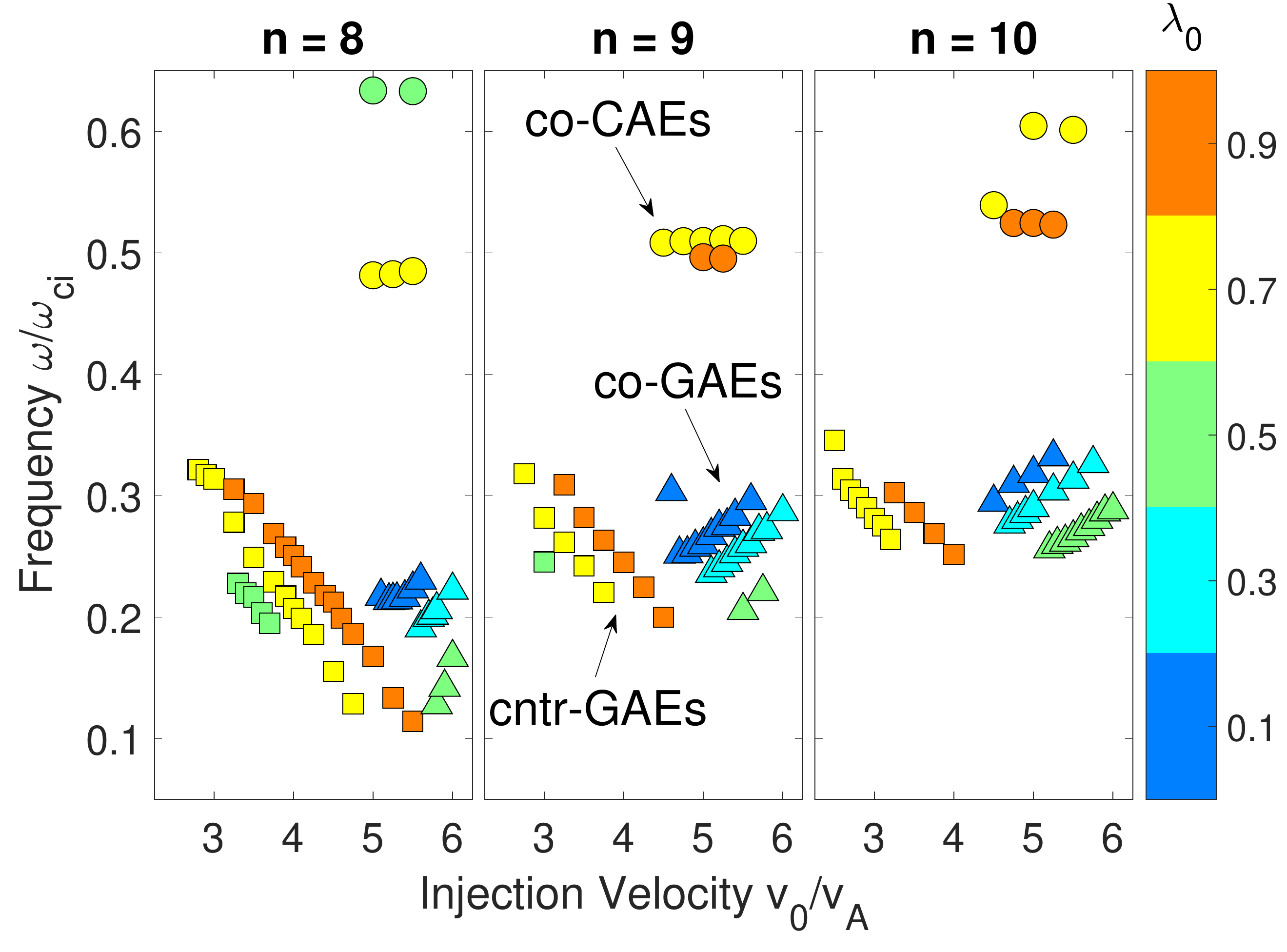}
\caption{Frequency of the most unstable modes for $\abs{n} = 8 - 10$ as a function of normalized injection velocity $\vinj$. Each plot shows modes for a single toroidal mode number $\abs{n}$, with cntr-GAEs marked by squares, co-GAEs marked by triangles, and co-CAEs by circles. Color denotes the central pitch $\linj$ of the fast ion distribution in each individual simulation.}
\label{fig:GAE_freq_shifts}
\end{figure}

The most interesting property of the GAEs in these simulations is how significantly the beam distribution influences the frequency of the most unstable mode. This discovery was thoroughly investigated in a recent publication \cite{Lestz2018POP}, and will be briefly summarized here. It was found that almost uniformly for each toroidal mode number, the frequency of the most unstable GAE scales linearly with the normalized injection velocity of the fast ion distribution. This effect is shown in \figref{fig:GAE_freq_shifts}, where the frequency of the most unstable mode can change by hundreds of kHz as $\vinj$ is varied in separate simulations. In most cases, the mode structure of the most unstable mode remains relatively stable despite these large changes in frequency -- yielding small quantitative changes in the mode's width, elongation, or radial location, but not changing mode numbers. This behavior is in contrast to what is seen for the CAEs, where the frequency of the most unstable mode is nearly unchanged for large intervals of $\vinj$, then switching to a new frequency at some critical value, which also coincided with the appearance of a new poloidal harmonic. In this sense, the GAEs behave more like energetic particle modes (EPMs), whose frequency fundamentally depends on fast ion properties, than conventional, perturbatively-driven MHD eigenmodes. Due to their sub-cyclotron frequency and $n < 0$, the cntr-GAEs can only interact with fast ions through the ordinary Doppler-shifted cyclotron resonance. It may be written as $\omega - n\avg{\omegator} - p\avg{\omegapol} = \avg{\omegaci}$.

\subsection{Co-propagating GAEs} 
\label{sec:gaep}

\begin{figure}[tb]
\includegraphics[width = \thirdwidth]{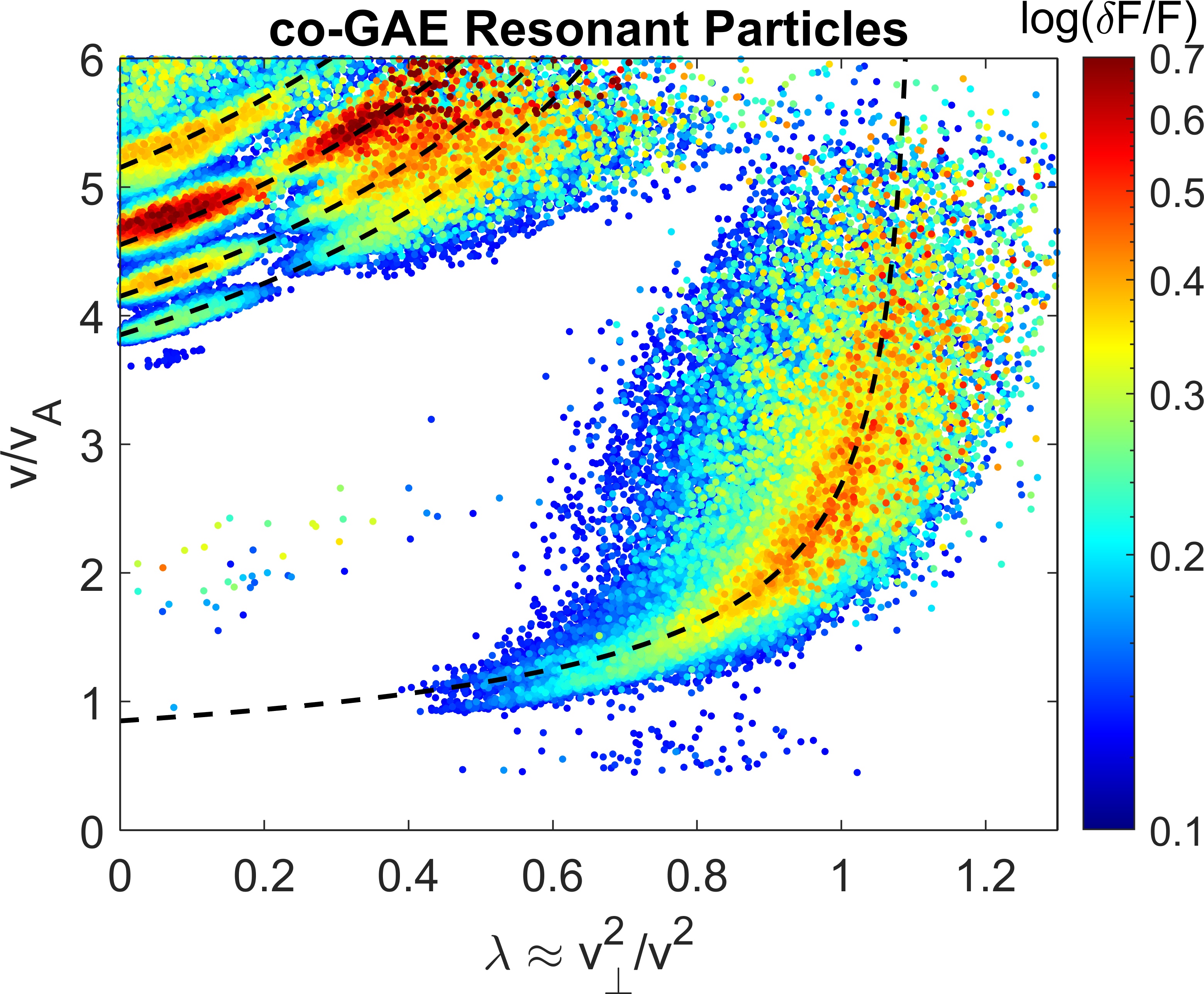}
\caption{Resonant particles for an $n = 9$ co-GAE driven by beam ions with $\linj = 0.3$ and $\vinj = 5.1$. Dashed lines indicate contours of constant $\vpres/\va = 0.85, 3.85, 4.15, 4.55, 5.15$, corresponding to resonances with $(\lres=0; p=0)$ and $(\lres = -1; p =1, 0, -1, -2)$, respectively. The color scale indicates the normalized particle weights $\delta f/\fbeam$ on a log scale.}
\label{fig:GAEPres}
\end{figure}

In addition to the counter-GAEs, high frequency co-propagating GAEs were also found to be unstable in these simulations with $0.15\omegaci < \omega < 0.60$ across $n = 8 - 12$. Almost uniformly these modes have $m = 0$ or 1, similar to the low $m$ cntr-GAEs shown in \figref{fig:GAEMn6l0.7v5.0} and \figref{fig:GAEMcomp}. Due to the large $n$ values and small $m$, co-GAEs tend to have large $\krat > 1$ in simulations. The co-GAEs may simultaneously resonate with the high energy fast ions through the ``anomalous'' Doppler-shifted cyclotron resonance, $\omega - n\avg{\omegator} - p\avg{\omegapol} = -\avg{\omegaci}$, as well as with fast ions with $\vpar\like\va$ through the Landau resonance $\omega - n\avg{\omegator} - p\avg{\omegapol} = 0$, as shown on \figref{fig:GAEPres}. Although there may be comparable numbers of particles participating in these two resonances in the simulations, the energy exchange is dominated by the high energy fast ions (interacting via the $\lres = -1$ resonance). The inclusion of realistic two-fluid effects may increase the efficiency of the Landau resonance for GAEs in experiments relative to what is present in the single fluid hybrid simulations \cite{Lestz2020p2}. Just as with the cntr-GAEs, the high frequency co-GAEs behave more like EPMs than MHD eigenmodes, exhibiting large changes in frequency in proportion to changes in $\vinj$, without any significant changes to the mode structure (also shown on \figref{fig:GAE_freq_shifts}). 

\begin{figure*}[htb]
\includegraphics[width = \fullwidth]{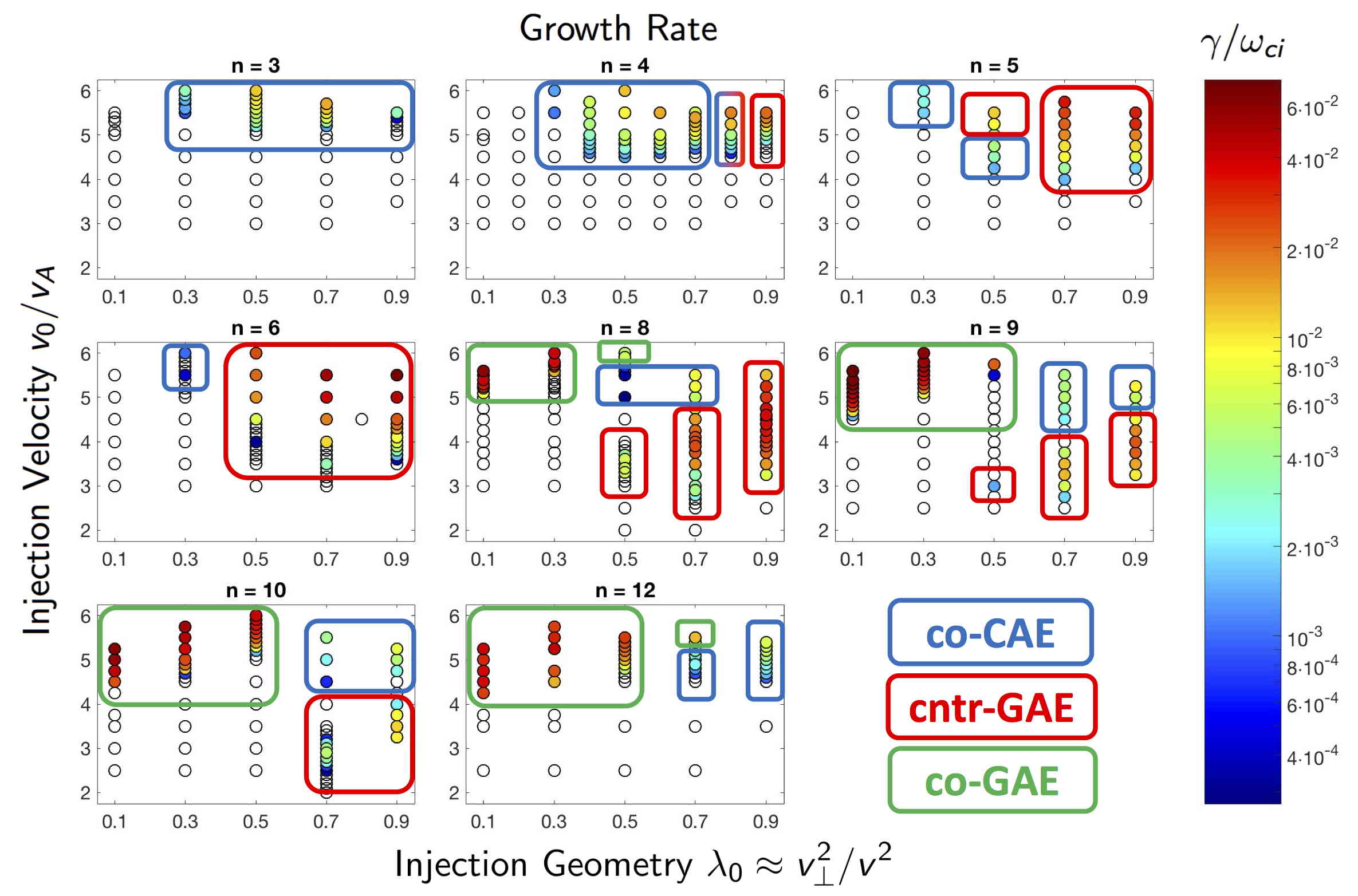}
\caption{Linear growth rates of CAE/GAE modes. Each subplot represents a single toroidal harmonic $|n|$, showing growth rate of the most unstable mode for each distribution parameterized by $(\linj,\vinj)$. Individual white data points represent simulations where no unstable mode was found. Colored circles indicate the magnitude of the growth rate. Data points enclosed by blue boxes are co-CAEs, red boxes are cntr-GAEs, and green boxes are co-GAEs.}
\label{fig:stab_all}
\end{figure*} 

Whereas cntr-GAEs are frequently observed experimentally and have been the subject of recent theoretical studies, high frequency co-GAEs are less commonly discussed. The early development of GAE theory did involve co-propagating modes, but these were restricted to very low frequencies and consequently only considered the $\lres = 0$ Landau resonance \cite{Ross1982PF,Appert1982PP,DeChambrier1982POP,Fu1989PF,VanDam1990FT}. This conventional type of low frequency co-GAE driven primarily by the Landau resonance was not found to be the most unstable mode for any set of fast ion parameters in these simulations. As explored in \citeref{Lestz2020p2}, this may be due to the fact that they have a similar instability condition to the co-CAEs, but with growth rate reduced by a typically small factor $(\omega/\omegaci)^2$. High frequency co-GAEs were never observed in NSTX, nor have they been documented in other devices. As detailed in \secref{sec:gammamaxlv}, this is at least partly because their resonance condition favors very tangentially injected, super-\Alfvenic beams which was not possible with the beam lines available on NSTX. However, the additional neutral beam installed on NSTX-U allows for much more tangential injection \cite{Gerhardt2012NF}, which should be able to excite high frequency co-GAEs for discharges with sufficiently large $\vinj$ (experimentally achievable with a low magnetic field). 

\section{Growth Rate Dependence on Beam Injection Geometry and Velocity} 
\label{sec:stabres}

\subsection{Interpretation of simulation results}
\label{sec:simres}

This section presents the linear stability trends from simulations, which will be subsequently interpreted with analytic theory. Unstable CAE/GAE modes are found for a variety of beam parameters, and all simulated toroidal mode numbers $|n| = 3 - 12$. In general, co-GAEs have the largest growth rate, followed by cntr-GAEs, and then co-CAEs. For realistic NSTX beam geometry $(\linj = 0.5 - 0.7)$, cntr-GAEs usually have the largest growth rate. 

The summary of results from a large set of simulations is shown in \figref{fig:stab_all}. Each subplot corresponds to simulations restricted to a single toroidal harmonic $\abs{n}$. Within each plot, each circle represents a separate simulation with the fast ion distribution in \eqref{eq:F0} parameterized by injection geometry $\linj \approx \vperp^2/v^2$ and injection velocity $\vinj$. The white circles indicate simulations where all modes were stable (a small random initial perturbation decays indefinitely), while the color scale indicates the growth rate of the most unstable mode. The most unstable mode in each simulation is identified as a co-CAE, cntr-GAE, or co-GAE based on the descriptions given in \secref{sec:id}. 

The normalized growth rates span three orders of magnitude in the simulations, ranging from $\gamma/\omegaci = 10^{-4}$ to $10^{-1}$. When instead normalized to the mode frequency, growth rates for CAEs are in the range $\gamma/\omega = 0.005 - 0.05$, while most of the GAEs have $\gamma/\omega = 0.01 - 0.2$, with a few more unstable cases having $\gamma/\omega$ up to 0.4. While there is reason to believe that GAE growth rates \emph{are} typically larger than those of CAEs, the nearly order of magnitude difference seen in these simulations is primarily due to the different resonant interactions, as will be explained shortly. The colored boxes on \figref{fig:stab_all} indicate the most unstable type of mode in each simulation: co-CAE, cntr-GAE, or co-GAE, as determined based on the properties discussed in \secref{sec:id}. 

\begin{figure}[tb]
\includegraphics[width = \thirdwidth]{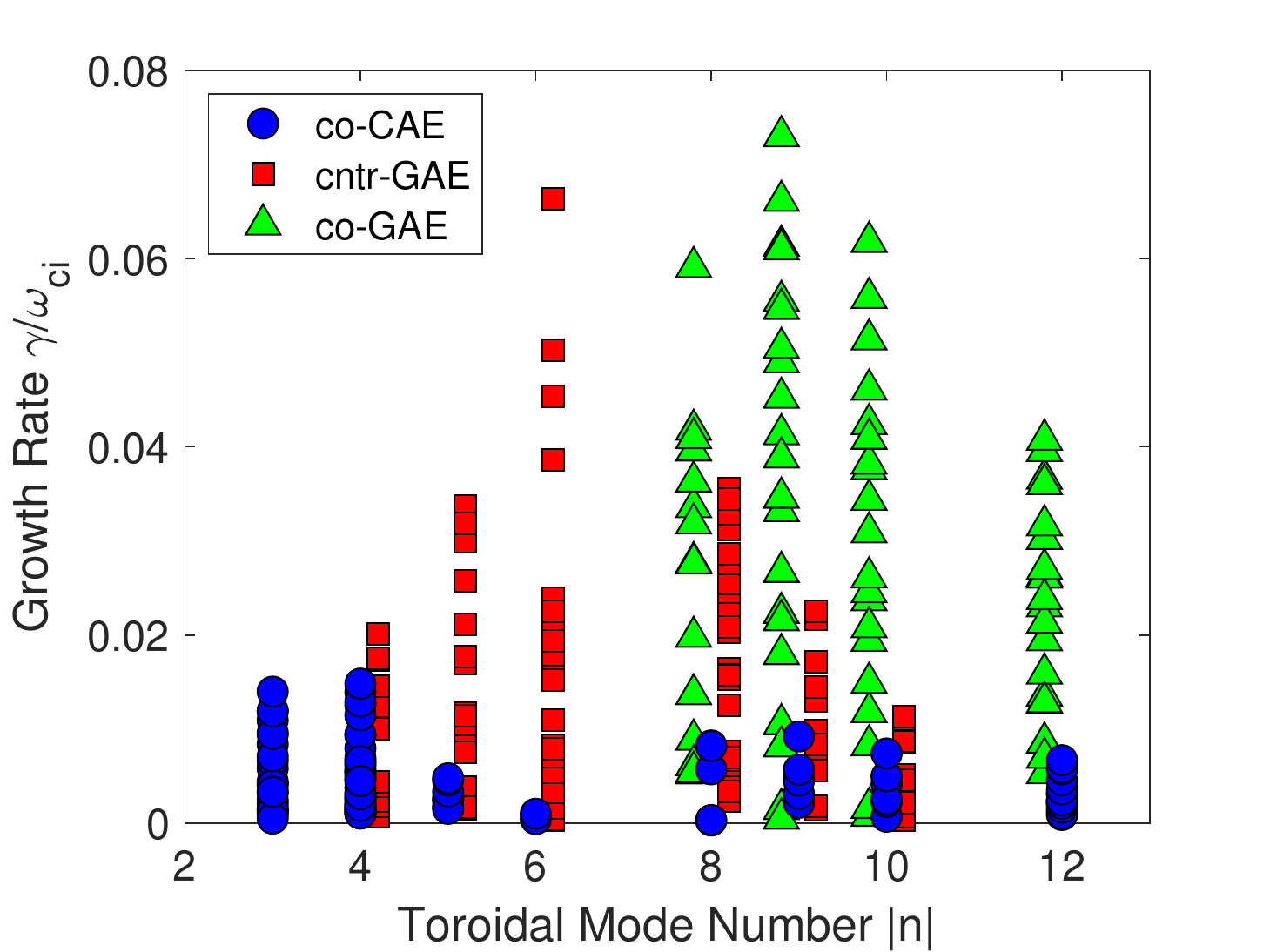}
\caption{Growth rate of each type of mode as a function of toroidal mode number in simulations.}
\label{fig:gamma_vs_n}
\end{figure}

Across all types of modes, the most unstable modes are found for $\abs{n} = 6 - 9$. The growth rate dependence on toroidal mode number is shown in\figref{fig:gamma_vs_n}. At low $\abs{n} = 3 - 4$, co-CAEs were the most common mode. For each value of $\abs{n} =  3 - 12$, there exists at least one set of beam parameters $(\linj,\vinj)$ such that a co-CAE was the most unstable mode, with the growth rate maximized for $n = 4$. In contrast, the GAE growth rates peaked at moderately large values of $\abs{n} = 6$ and $\abs{n} = 9$ for cntr-GAEs and co-GAEs, respectively. Note that the co-GAEs are excited only at large $n$ since the $\lres = -1$ resonance requires a large Doppler shift $\kpar\vpres \approx n\vpres/R$. In simulations restricted to $\abs{n} = 1$ and $\abs{n} = 2$, the only unstable modes had $\omeganorm < 0.05$, with high $m$ numbers and mixed polarization. Hence these modes are qualitatively different from the CAEs/GAEs being studied, and their further investigation is left to future work. 

The simulation results can be interpreted with linear analytic theory. In \citeref{Lestz2020p1} and \citeref{Lestz2020p2}, a local expression for the fast ion drive due to an anisotropic neutral beam-like distribution is derived. Notably, the effect of finite injection energy of NBI distributions is included consistently, yielding a previously overlooked instability regime \cite{Belova2019POP}. Terms to all orders in $\omeganorm$, $\krat$, and $\kperp\rhob$ are kept for applicability to the entire possible spectrum of modes. Analysis was restricted to 2D velocity space, neglecting the influence of plasma non-uniformities. In particular, this excludes drive/damping due to gradients in $\pphi$, which is addressed in \secref{sec:pphi}. Moreover, the calculation does not include bulk damping sources, hence it is most reliable when applied far from marginal stability. To supplement this analysis, quantification of the magnitude of damping present in \HYM simulations as well as an estimate of the thermal electron damping rate (absent in the simulation model) will be presented in \secref{sec:damp}. In a nutshell, the drive (or damping) due to fast ions is a weighted integral over gradients of the fast ion distribution:

\begin{align}
\gamma \appropto \int d\lambda h(\lambda)\left[\left(\frac{\lres\omegaci}{\omega} - \lambda\right)\pderiv{}{\lambda} + \frac{v}{2}\pderiv{}{v}\right]\fb
\label{eq:gamma_dfdxdv}
\end{align} 

Here $\lres$ is the cyclotron resonance coefficient appearing in \eqref{eq:reskpar} and $h(\lambda)$ is a complicated positive function that weights the integrand. The full expression for the fast ion drive determined by \eqref{eq:F0} can be found in Eq. 21 of \citeref{Lestz2020p1}, which can be integrated numerically for given beam and mode parameters in order to quantitatively predict the fast ion drive. However, we can first gain a qualitative understanding of the simulation results by analyzing \eqref{eq:gamma_dfdxdv}. For the model distribution, the $\partial\fb/\partial v$ term is always negative (damping) since the slowing down function decreases monotonically. Velocity space anisotropy $(\partial\fb/\partial\lambda)$ can provide either drive or damping depending on its sign. For $\lres \neq 0$ resonances and the experimental value of $\dl \approx 0.3$, the $\partial\fb/\partial v$ contribution is much smaller than that from $\partial\fb/\partial\lambda$, though they can be more comparable when $\lres = 0$.

\begin{figure}[tb]
\includegraphics[width = \thirdwidth]{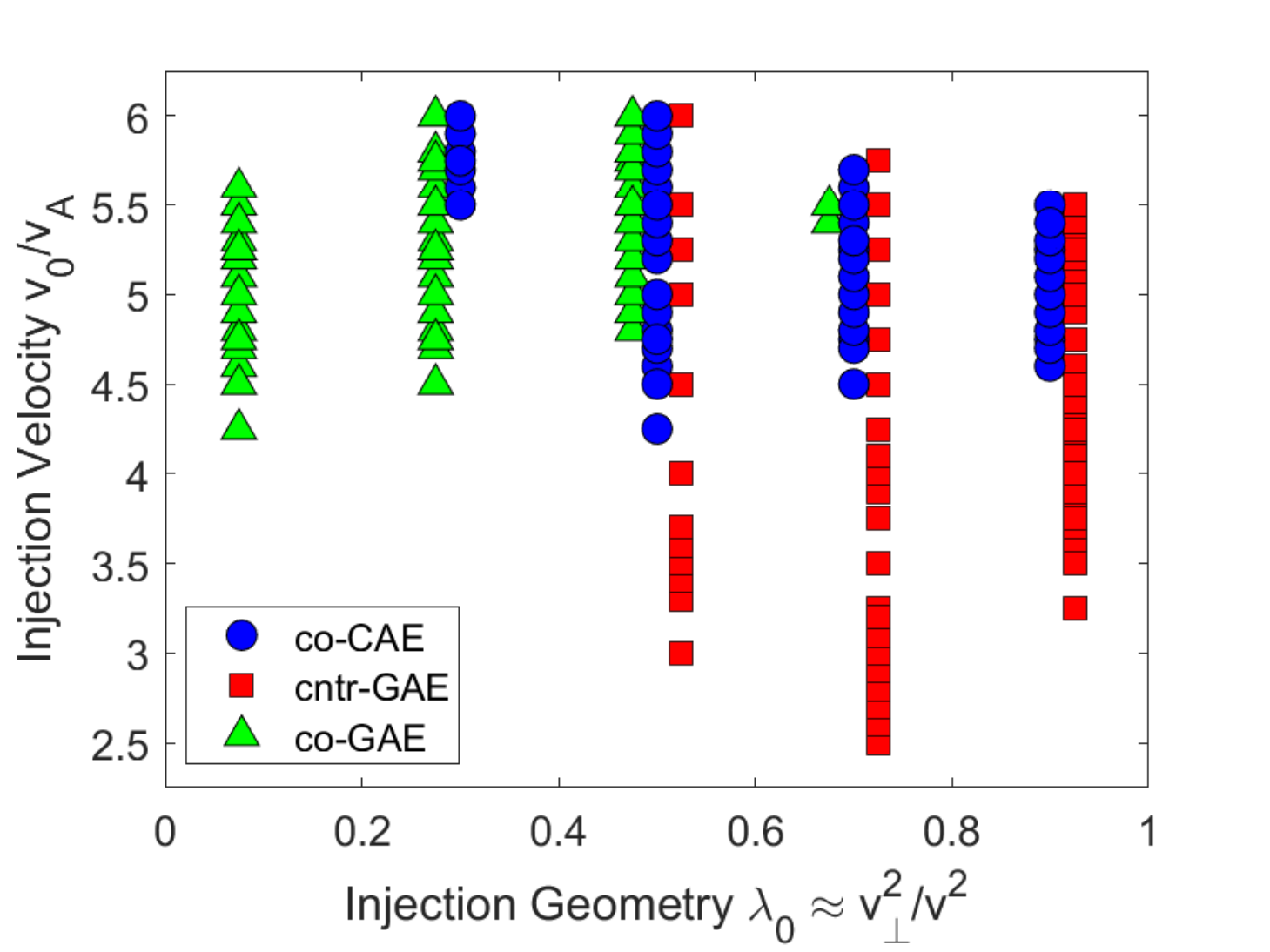}
\caption{Existence of unstable mode type as a function of the beam injection geometry $\linj$ and velocity $\vinj$ in simulations.}
\label{fig:lv_prop}
\end{figure}

Now it is clear why the co-CAE growth rates are typically smaller than those of the GAEs. As discussed previously, the co-CAEs are driven by the $\lres = 0$ resonance, while the co-GAEs and cntr-GAEs are driven by $\lres = \pm 1$, which leads to the large factor $\omegaci/\omega$ multiplying the GAE growth rates. Cntr-CAEs have also been observed in experiments \cite{Crocker2013NF,Sharapov2014POP}, which should also have relatively large growth rates due to this factor for the $\lres = 1$ resonance, but they are not seen as the most unstable modes in \HYM simulations. As discussed in Sec. V of \citeref{Lestz2020p1}, local theory predicts the linear growth rate of cntr-CAEs to be somewhat smaller than that of cntr-GAEs driven by the same fast ion distribution, so these subdominant modes would not appear in linear initial value simulations. 

One might also ask why the co-GAEs are driven by the $\lres = -1$ resonance, but the co-CAEs are not, since it would enhance their growth rate based on the argument above. Due to the difference in dispersion, fast ions must have a larger parallel velocity to resonate with CAEs than GAEs. For GAEs, the $\lres = -1$ resonance requires $\vpres/\va \approx (1 + \omegaci/\omega)$, while for CAEs, $\vpres/\va \approx \abs{k/\kpar}(1 + \omegaci/\omega)$. For the ranges of $\abs{k/\kpar}$ and $\omegaci/\omega$ inferred from modes in the simulations, this resonance would require fast ion velocities far above the simulated beam injection velocity, hence the condition can not be satisfied and co-CAEs are restricted to the $\lres = 0$ resonance with typically smaller growth rates. 

In order to compare the stability properties of each mode against one another as a function of beam parameters, the results contained in \figref{fig:stab_all} have been condensed into \figref{fig:lv_prop}, which includes all simulated toroidal harmonics and examines the existence of modes instead of the magnitude of their growth rate. Clearly, the cntr-GAEs prefer large $\linj$ whereas the co-GAEs prefer small $\linj$. This is reasonable  based on the form of \eqref{eq:gamma_dfdxdv}. Cntr-GAEs interacting with beams ions through the $\lres = 1$ resonance are driven by $\partial\fb/\partial\lambda > 0$, while co-GAEs are driven by regions of phase space with $\partial\fb/\partial\lambda < 0$ due to the $\lres = -1$ resonance. Hence when the distribution peaks at large $\linj \rightarrow 1$, a larger region of phase space contributes to drive the cntr-GAEs (and damp the co-GAEs). Conversely when $\linj \rightarrow 0$, more resonant fast ions contribute to driving the co-GAEs (and damp cntr-GAEs). A more quantitative comparison will be shown in \secref{sec:gammamaxlv} to elaborate on this intuition.  

The co-CAE dependence on $\linj$ is somewhat more subtle. From \eqref{eq:gamma_dfdxdv}, one would expect that small $\linj$ favors their excitation, just as in the previous argument given for co-GAEs. Yet, \figref{fig:lv_prop} shows unstable CAEs across a wide range $0.3 \leq \linj \leq 0.9$. Although the fast ions do provide drive for co-CAEs for $\linj = 0.1$ according to theory, the predicted growth rate is too small to overcome the background damping in simulations. As will be shown in \secref{sec:gammamaxlv}, numerical evaluation of \eqref{eq:gamma_dfdxdv} predicts that the drive from fast ions peaks at some intermediate injection geometry $\linj \approx 0.5$, which is similar to what occurs in simulations which also include damping on the background plasma. 

\newcommand{\lvcomplen}{\halfwidth}
\begin{figure}[H]
\subfloat[\label{fig:lv_comp:caez}]{\includegraphics[width = \lvcomplen]{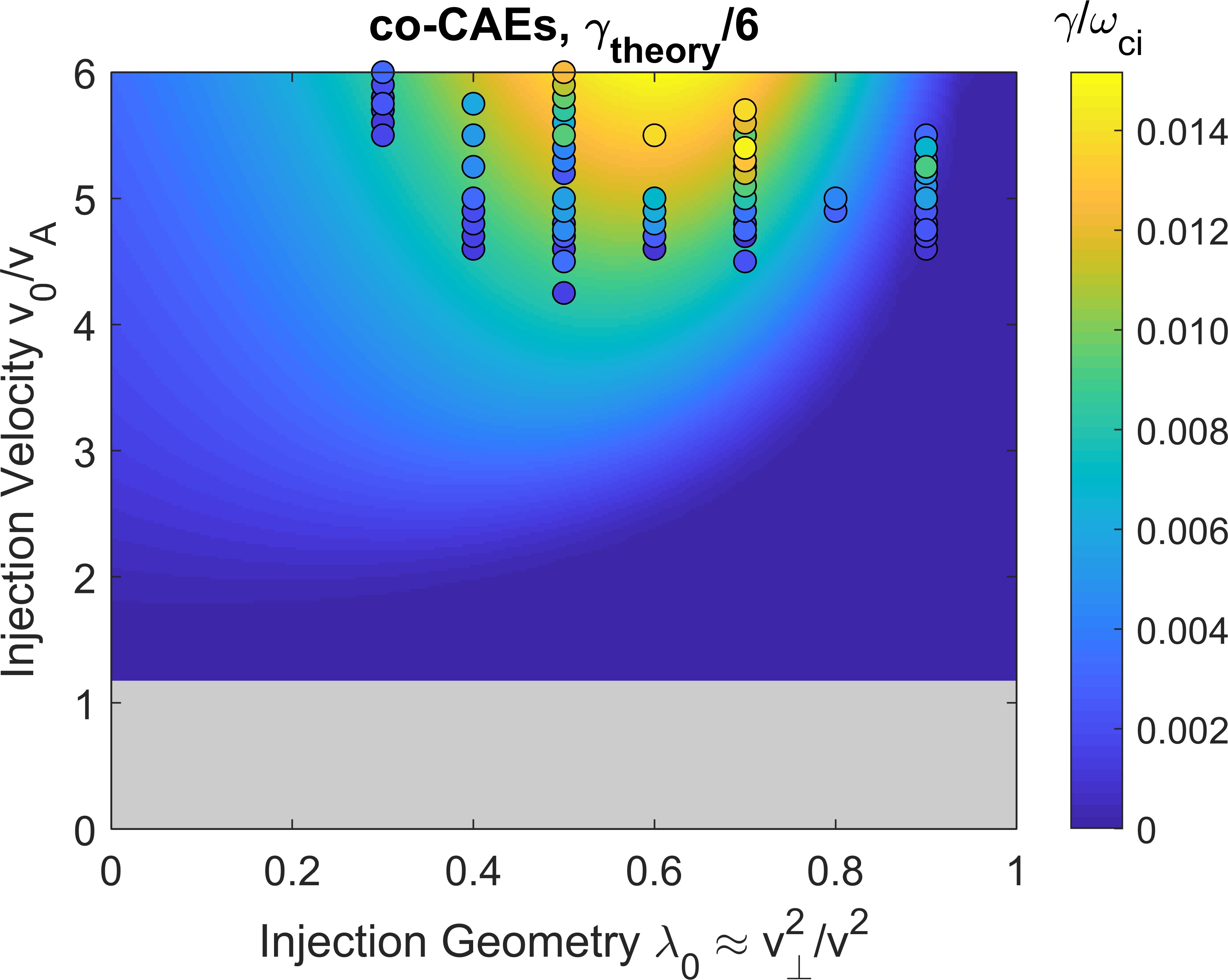}} \\
\subfloat[\label{fig:lv_comp:gaem}]{\includegraphics[width = \lvcomplen]{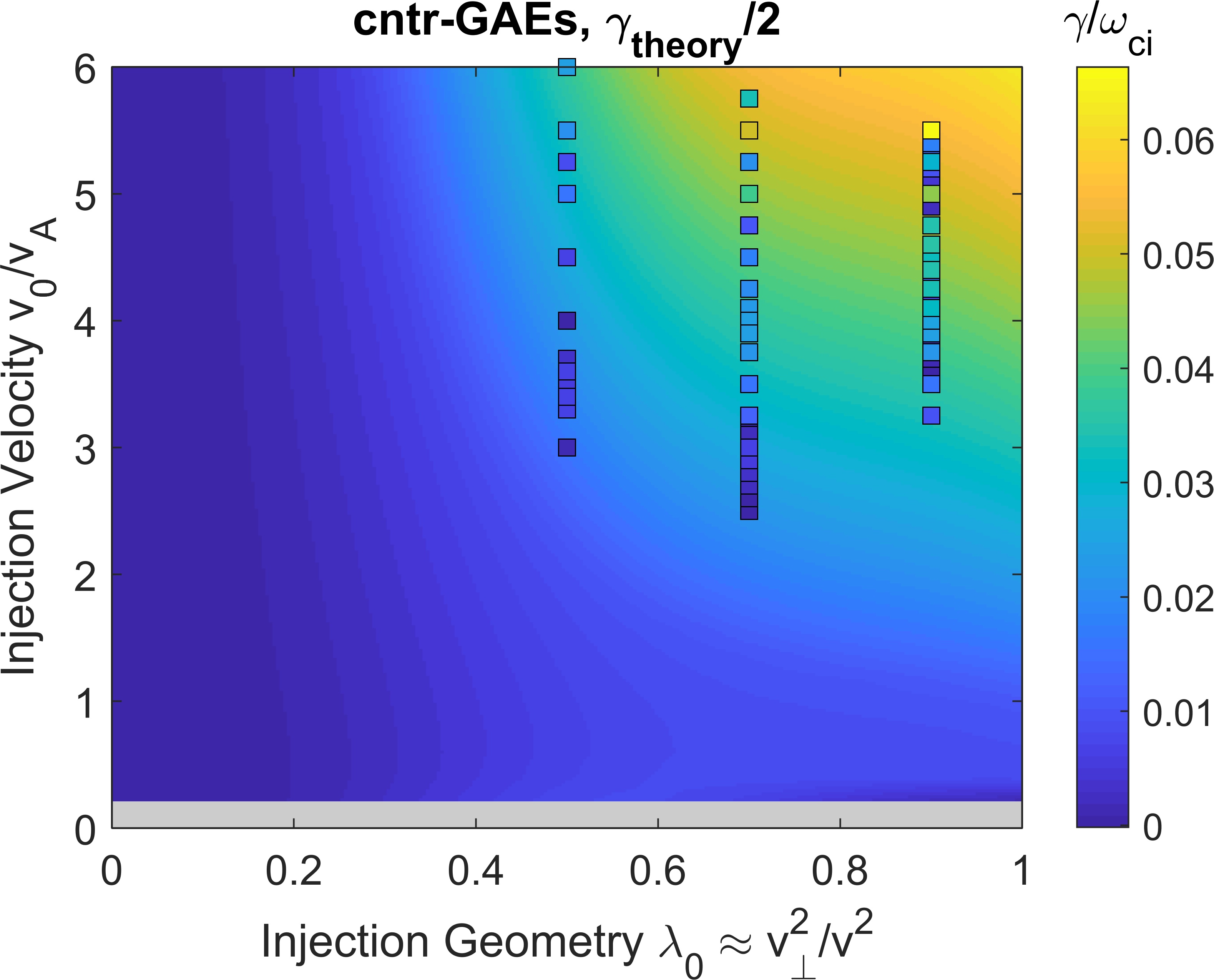}} \\
\subfloat[\label{fig:lv_comp:gaep}]{\includegraphics[width = \lvcomplen]{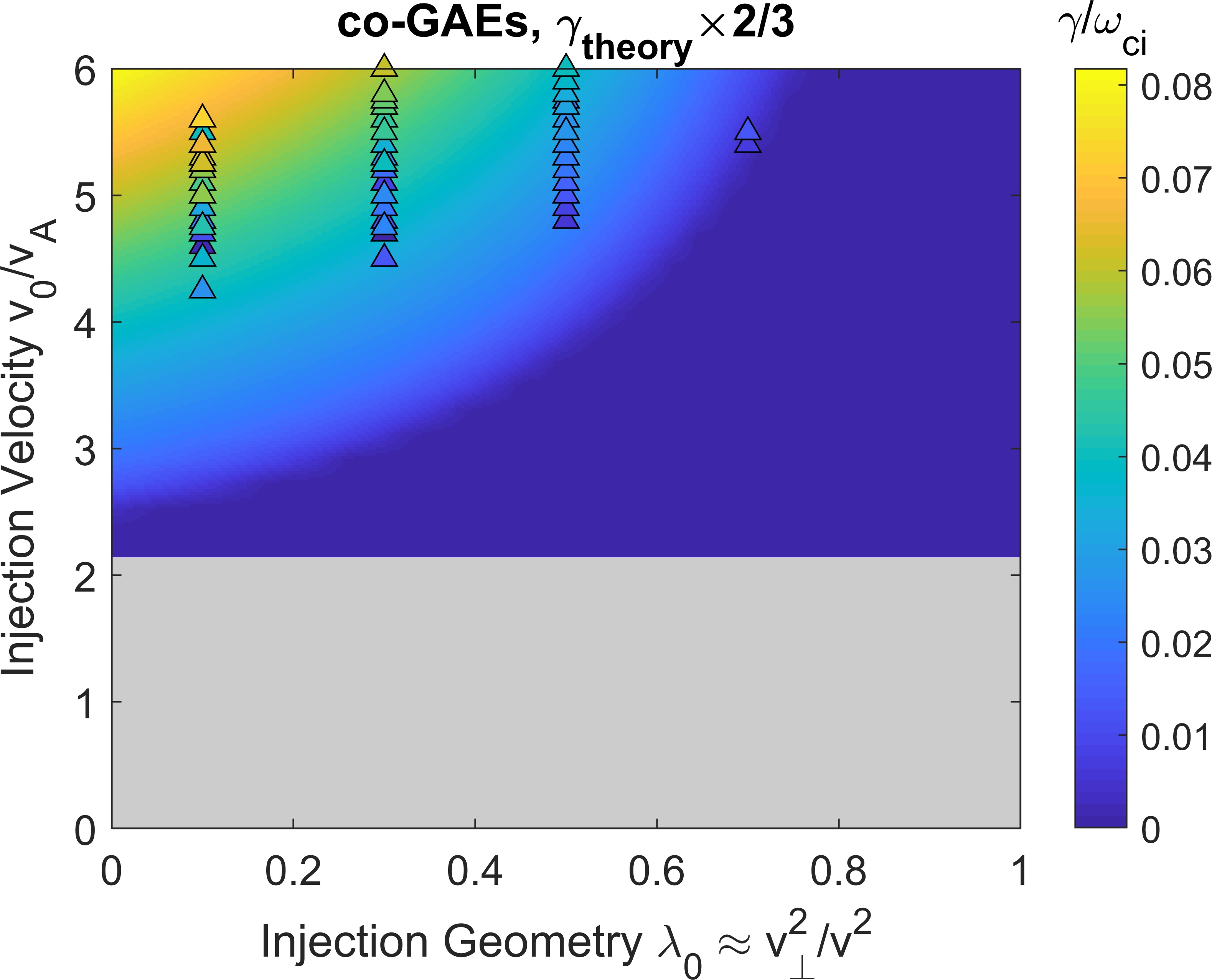}}
\caption{Growth rate of most unstable mode as a function of injection geometry $(\linj)$ and velocity $(\vinj)$, calculated from analytic theory (background color) and \HYM simulations (individual points) for (a) co-CAEs, (b) cntr-GAEs, and (c) co-GAEs. In order to more clearly show trends, the analytically calculated growth rate has been re-scaled as indicated.}
\label{fig:lv_comp}
\end{figure}

With respect to the injection energy, the cntr-GAEs are unstable at significantly lower beam voltages than the CAEs. Whereas no CAE is found to be unstable for $\vinj < 4$, unstable cntr-GAEs were found with $\vinj > 2.5$. This is consistent with NSTX experiments, where cntr-GAEs were more routinely observed and in a wider array of operating parameters. This result is also qualitatively similar to dedicated MAST experiments performed at a range of magnetic field values $(\vinj = 1.5 - 2.25)$ which found a transition from cntr- to co-propagating modes as $\vinj$ was increased \cite{Sharapov2014POP}. For co-CAEs driven by the $\lres = 0$ resonance, the fast ion damping from $\partial\fb/\partial v$ competes more closely with the drive from anisotropy $\partial\fb/\partial\lambda$ than when $\lres \neq 0$, leading to a larger $\vinj$ for instability. As discussed in Sec. III.B.1 of \citeref{Lestz2020p2}, net drive for co-CAEs driven by beams with $\linj = 0.7$ and $\dl = 0.3$ requires $\vinj > 4.1$, similar to what is found in \HYM simulations. The co-GAEs also require relatively large $\vinj$ for excitation, due to the requirement of a sufficiently large Doppler shift $\kpar\vpres$ in order to satisfy the strong $\lres = -1$ resonance which drives them. Note that $\vinj \gtrsim 2.5$ and $\vinj \gtrsim 4$ should not be regarded as universal conditions necessary for cntr-GAE and co-GAE excitation, respectively. CAE/GAE excitation also depends on equilibrium profiles, which determine the background damping rate as well as the spectrum of eigenmodes. For instance, cntr-GAEs were routinely excited in early operation of NSTX-U \cite{Fredrickson2018NF} with $\vinj \approx 1 - 2$, while co-CAEs were rarely observed. In addition, cntr-propagating, sub-cyclotron modes have also been observed in dedicated low field DIII-D discharges \cite{Heidbrink2006NF,Tang2021U} with $\vinj \approx 1$, in agreement with corresponding \HYM simulations \cite{Belova2020IAEA}.

\subsection{Comparison of growth rate trends with theory}
\label{sec:gammamaxlv}

In this section, we will quantitatively examine the dependence of the growth rate on the beam injection geometry $\linj$ and velocity $\vinj$. Since the growth rate is sensitive to the specific mode properties (frequency $\omeganorm$ and wave vector direction $\krat$) whereas only the mode with the largest growth rate survives in the simulations, it makes sense to theoretically calculate the growth rate of the most unstable mode across all values of $0 < \omeganorm < 1$ and $\krat$, using the growth rate expression in Eq. 21 of \citeref{Lestz2020p1}. While that derivation contains general two fluid effects, they will be neglected here since our aim is interpretation of the simulations which use a single fluid background plasma. The growth rate dependence is shown in \figref{fig:lv_comp}, with separate subplots for co-CAEs, cntr-GAEs, and co-GAEs. In these figures, the background color gives the growth rate calculated analytically, while the individual points show the growth rates of unstable modes in separate simulations. 

The analytic growth rate is typically larger than that from simulations, so their magnitudes have been re-scaled as indicated in the figures so that a comparison of trends can be made. It is unsurprising that the analytic growth rates are larger than those from simulations for two reasons. First, the simulations include drive/damping from fast ions and also damping of the mode on the single fluid resistive background plasma. As will be discussed in \secref{sec:damp}, this damping can be of order $20\% - 60\%$ of the fast ion drive, while it is not present at all in the analytic calculation which perturbatively considers the contribution from fast ions alone. Second, the analytic theory being used is a \emph{local} approximation which does not take into account equilibrium profiles of the background plasma, fast ion density, or spatial dependence of the mode structure. The value of $\nb$ that we use in the local calculation is its peak value ($5.3\%$ for the simulations in this section), leading to a calculated growth rate larger than it would be if spatial dependencies were taken into account. Notably, the local calculation does not include the contribution of $\partial\fb/\partial\pphi$ to the growth rate, which can be either positive or negative depending on the sign of $n$, as will be discussed. For the high frequency modes studied in this work, this contribution is typically smaller than the terms which are kept in the local approximation, but it nonetheless represents a potential source of error in this simplified calculation of the growth rate. Consequently, the value of comparing this calculation with the simulation results lies in the trends with beam parameters, not the absolute values of the growth rate.

We see that for each type of mode, there is very reasonable agreement between the regions of instability predicted by analytic theory and the beam parameters of unstable modes. For co-CAEs, theory predicts the largest growth rate near moderate values of $\linj \approx 0.5$ and large $\vinj$, which are also the beam parameters preferred by unstable modes in simulations. According to theory, cntr-GAEs generally become more unstable for larger values of $\linj$ (more perpendicularly injected beams), while co-GAEs are most unstable for small $\linj$ (very tangential injection). This is exactly the trend seen in simulations, where the unstable cntr-GAEs generally have largest growth rate for  $\linj = 0.7 - 0.9$ and the co-GAEs are most unstable for $\linj = 0.1$ (the smallest simulated value). Hence, numerical evaluation of the analytic expression for fast ion drive confirms many of the qualitative arguments given when interpreting \figref{fig:lv_comp} through the lens of \eqref{eq:gamma_dfdxdv} in \secref{sec:simres}. For all types of modes, larger beam velocity leads to a larger maximum growth rate, though that growth rate may correspond to different mode parameters $\omeganorm$ and $\krat$ than the most unstable mode for some smaller beam velocity. 

Lastly, note that the trends from the calculation shown in \figref{fig:lv_comp} rely on the aforementioned search over all mode parameters for the most unstable mode. For instance, when the same calculation is done for cntr-GAEs with fixed values of $\omeganorm$ and $\krat$, the growth rate is no longer a strictly increasing function of $\vinj$ and $\linj$, but rather it can peak and then decrease, as in Fig. 2 of \citeref{Lestz2020p1}. This occurs when the beam parameters are varied in such a way that not as many particles resonate with the particular mode of interest. An example of such behavior is given in the $\abs{n} = 8$ subplot of \figref{fig:stab_all}, as the cntr-GAE growth rate for injection geometry $\linj = 0.7$ first increases, peaks, and then decreases, with the cntr-GAE eventually being replaced by a more unstable co-CAE. Since only the $\abs{n} = 8$ toroidal harmonic is kept in that simulation, the range of unstable modes of a given type is also restricted.  

\subsection{Approximate stability boundaries}
\label{sec:stabapprox}

Approximate stability boundaries with respect to beam injection parameters for CAEs and GAEs have recently been derived under conditions relevant to the simulated NSTX conditions \cite{Lestz2020p1,Lestz2020p2}. Such boundaries apply when $0.2 \lesssim \dl \lesssim 0.8$ (the NSTX(-U) value is $\dl \approx 0.3$) and either $\zp \defined \kperp\vpres/\omegaci \lesssim 2$ or $\zp \gg 2$. Note that this constraint is related to a condition on the finite Larmor radius (FLR) of the fast ions, since $\kperp\rhob \approx \zp\sqrt{\lambda / (1 - \lambda)}$. Moreover, the parameter $\zp$ can be re-written using the resonance condition as $\zp = \abs{\omega - \lres\avg{\omegaci}}\kratinv/\omegaci$, allowing the calculation of $\zp$ from mode properties alone. Using the data shown in \figref{fig:ao_prop}, one can calculate that $\zp = 0.5 - 1$ for the co-CAEs and $\zp = 0.5 - 1.5$ for the co-GAEs, so all of the simulated co-propagating modes fall within the $\zp \lesssim 2$ regime. Meanwhile, $\zp = 0.5 - 3$ for cntr-GAEs, though the most unstable ones do have $\zp < 2$. Hence the approximate instability conditions derived in Sec. IV.B.1 of \citeref{Lestz2020p1} for $\lres = \pm 1$ resonances and Sec. III.B.1 of \citeref{Lestz2020p2} for the $\lres = 0$ resonance in the small FLR regime are applicable to the unstable modes in NSTX simulations presented here. Using these approximations, and defining $\omegacires \defined \avg{\omegaci}/\omegaci$ for convenience, the following condition is found for GAE instability due to beam anisotropy 

\begin{align}
\label{eq:cntrGAE}
\vb &< \frac{\vpres}{(1 - \linj\omegacires)^{3/4}} \eqlab{cntr-GAEs} \\ 
\label{eq:coGAE}
\vb &> \frac{\vpres}{(1 - \linj\omegacires)^{3/4}} \eqlab{co-GAEs} 
\end{align}

Meanwhile, for co-CAEs, the $\partial\fb/\partial v$ terms must also be taken into account, which leads to a more complicated condition (where we also define $\xinj = \omegacires\linj$ and $\dx = \omegacires\dl$ to shorten the expression)

\begin{align}
\label{eq:coCAE}
\vb &> \vpres\left[\frac{1 - \dx\sqrt{2/3}}{\left[1 - \frac{1}{2}\left(\xinj + \sqrt{\xinj^2 + 8\dx^2/3}\right)\right]
\left[1 - \dx^{4/5}\right]}\right]^{5/8}
\end{align}

\begin{figure}[tb]
\subfloat[\label{fig:quant_gae}]{\includegraphics[width = \thirdwidth]{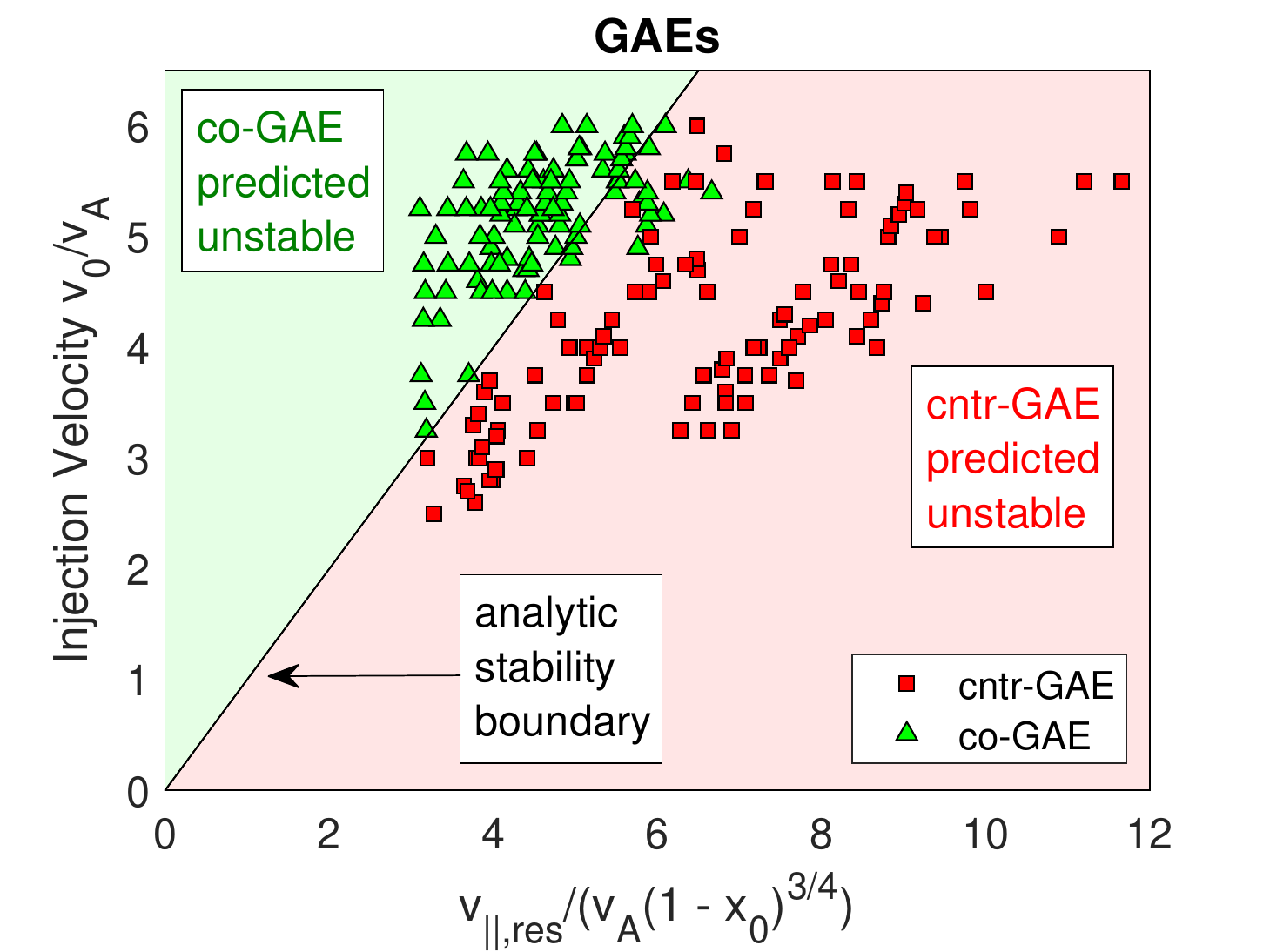}} \\
\subfloat[\label{fig:quant_cae}]{\includegraphics[width = \thirdwidth]{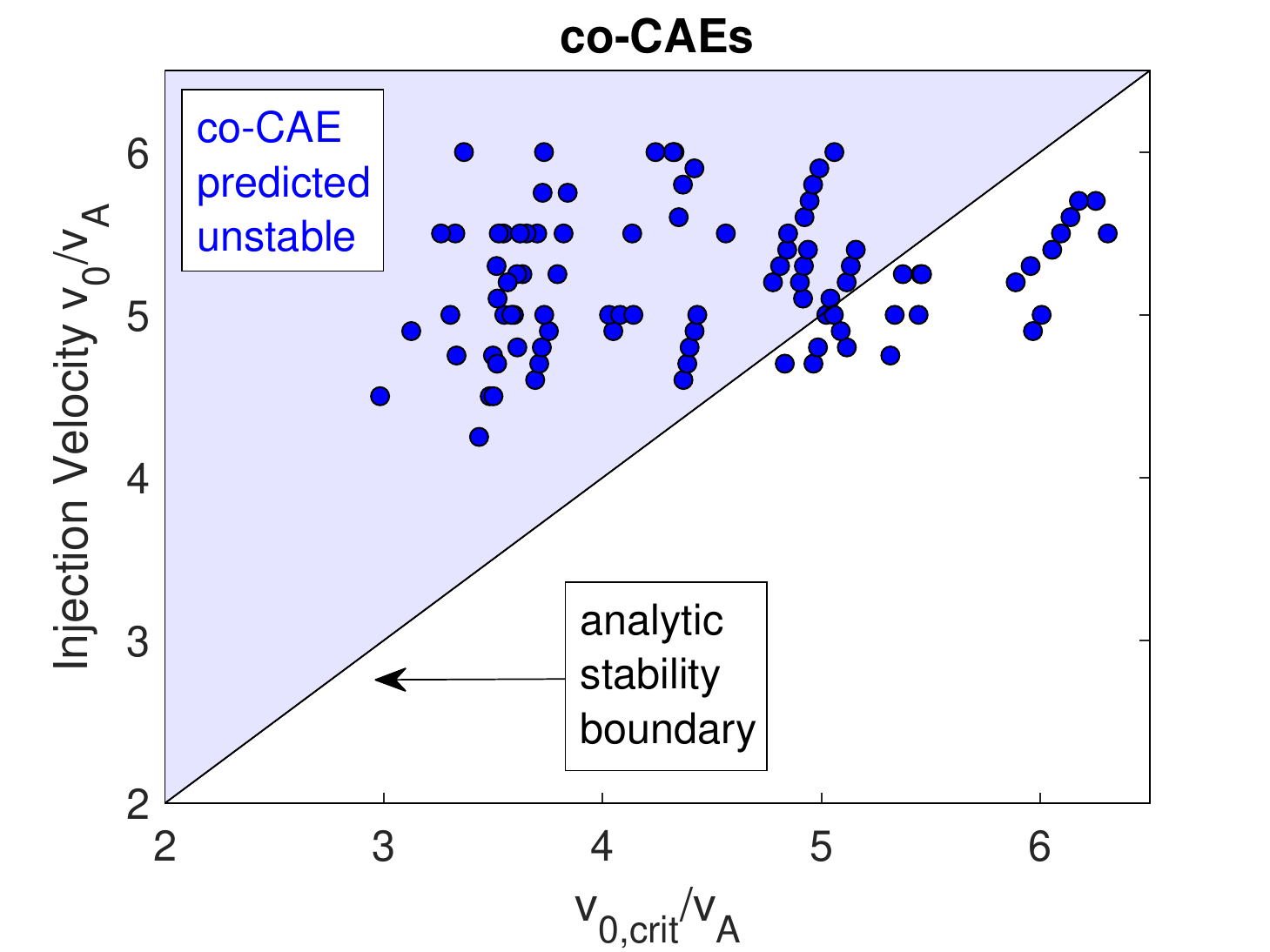}} 
\caption{Quantitative comparison of unstable modes in simulations against analytic predictions for (a) GAEs and (b) CAEs. The shaded areas indicate regions of instability predicted by the approximate conditions given in \eqref{eq:cntrGAE}, \eqref{eq:coGAE}, and \eqref{eq:coCAE} for cntr-GAEs, co-GAEs, and co-CAEs, respectively. The points are unstable modes of the labeled types from simulations. In (b), $v_{0,\text{crit}}$ is defined as the right hand side of \eqref{eq:coCAE}}
\label{fig:quant}
\end{figure}

Note that in each of the above equations, $\vpres$ implicitly depends on $\omeganorm$ and $\krat$ through the resonance condition and the appropriate dispersion relation for each mode. Hence, these conditions place constraints on the beam parameters ($\linj$, $\vinj$, as well as $\dl$ for co-CAEs) and mode properties for a mode to be driven unstable by a neutral beam distribution. Similar to our previous analysis, \eqref{eq:cntrGAE} and \eqref{eq:coGAE} demonstrate that the cntr-GAEs and co-GAEs have complementary instability conditions resulting from the opposite sign of $\lres$ in the resonance driving each mode. Consequently, co-GAE excitation is favored when $\linj$ is small, while cntr-GAEs prefer large $\linj$. 

\newcommand{\aocomplen}{\halfwidth}
\begin{figure}[H]
\subfloat[\label{fig:ao_comp_caez}]{\includegraphics[width = \aocomplen]{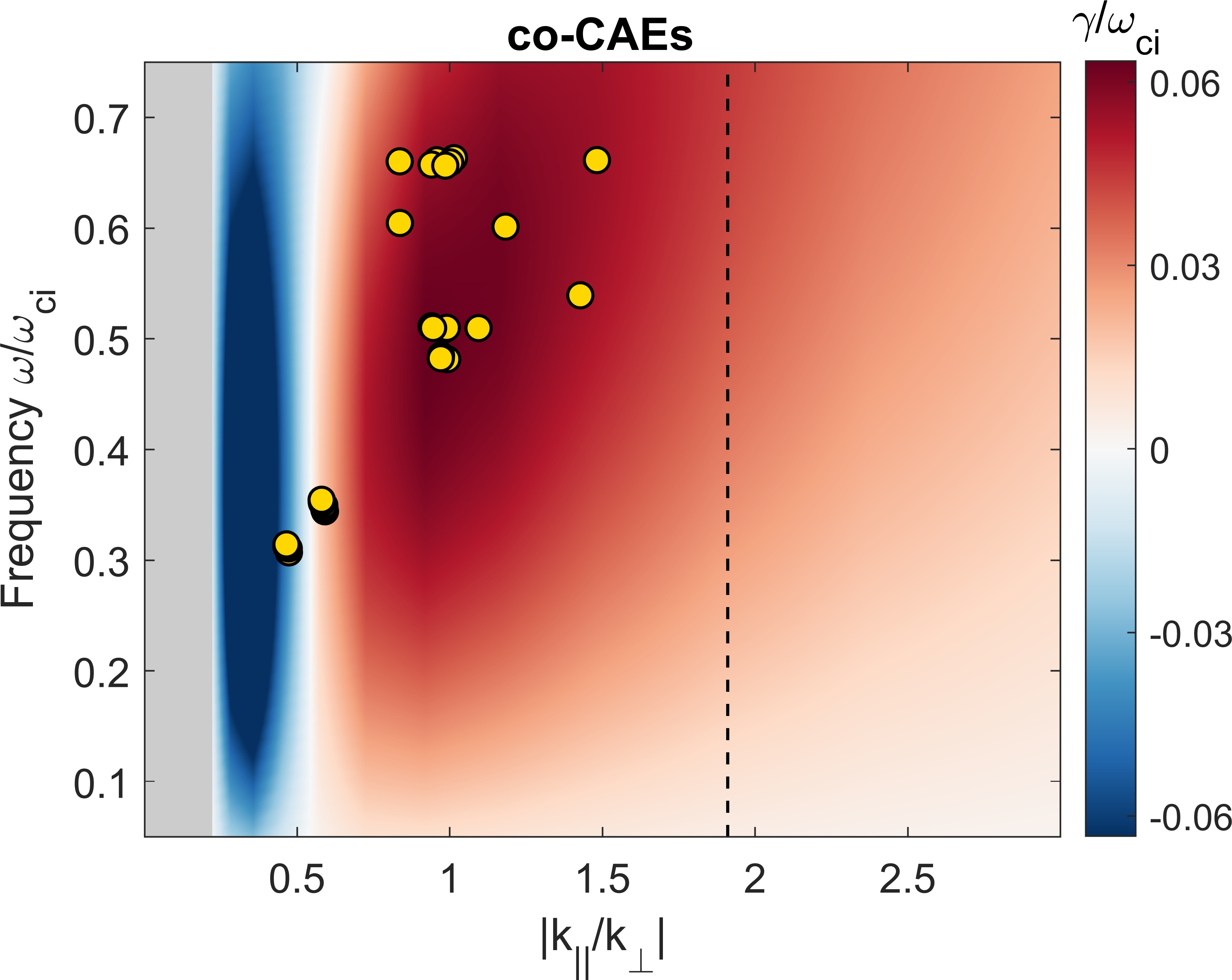}} \\
\subfloat[\label{fig:ao_comp_gaem}]{\includegraphics[width = \aocomplen]{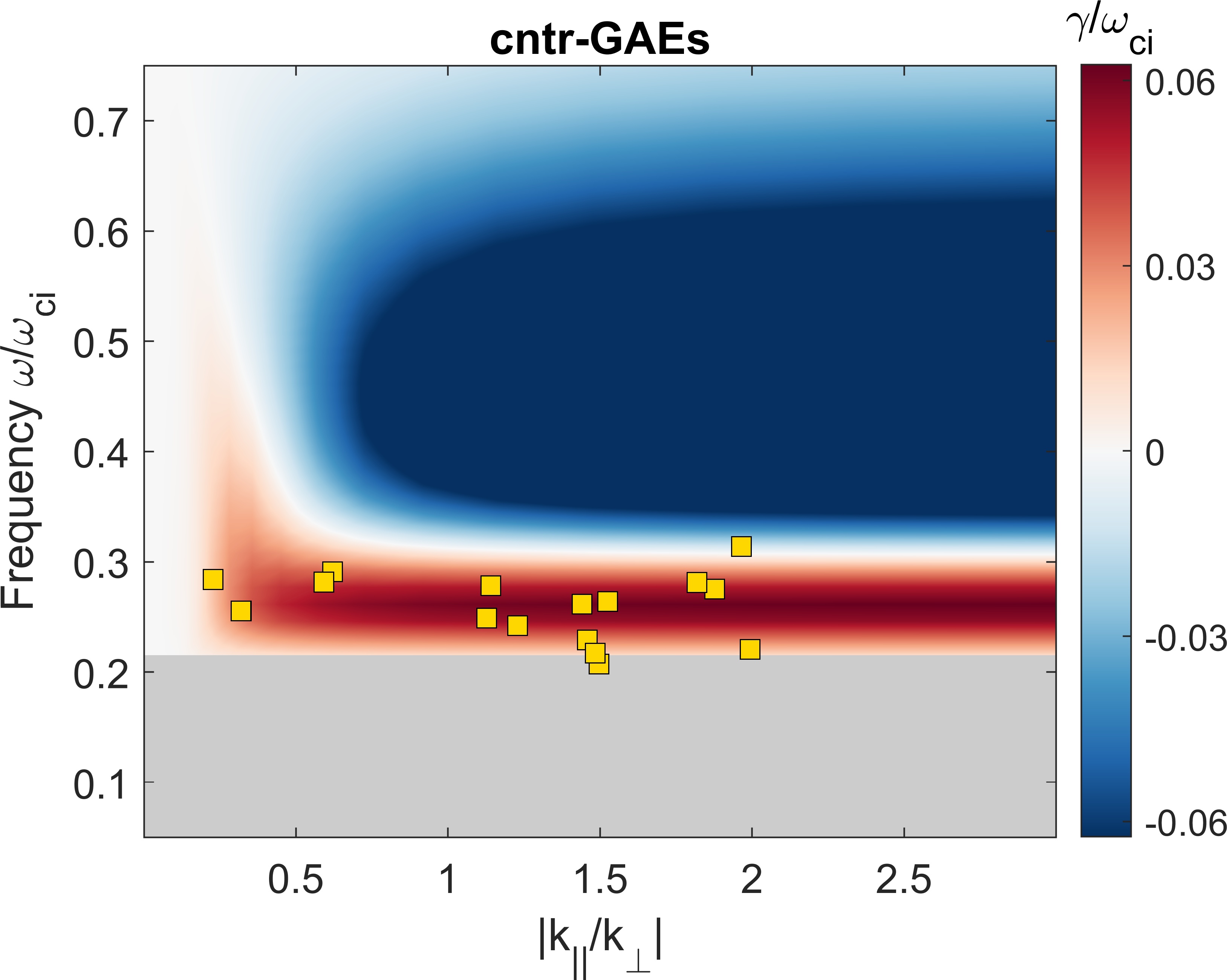}} \\
\subfloat[\label{fig:ao_comp_gaep}]{\includegraphics[width = \aocomplen]{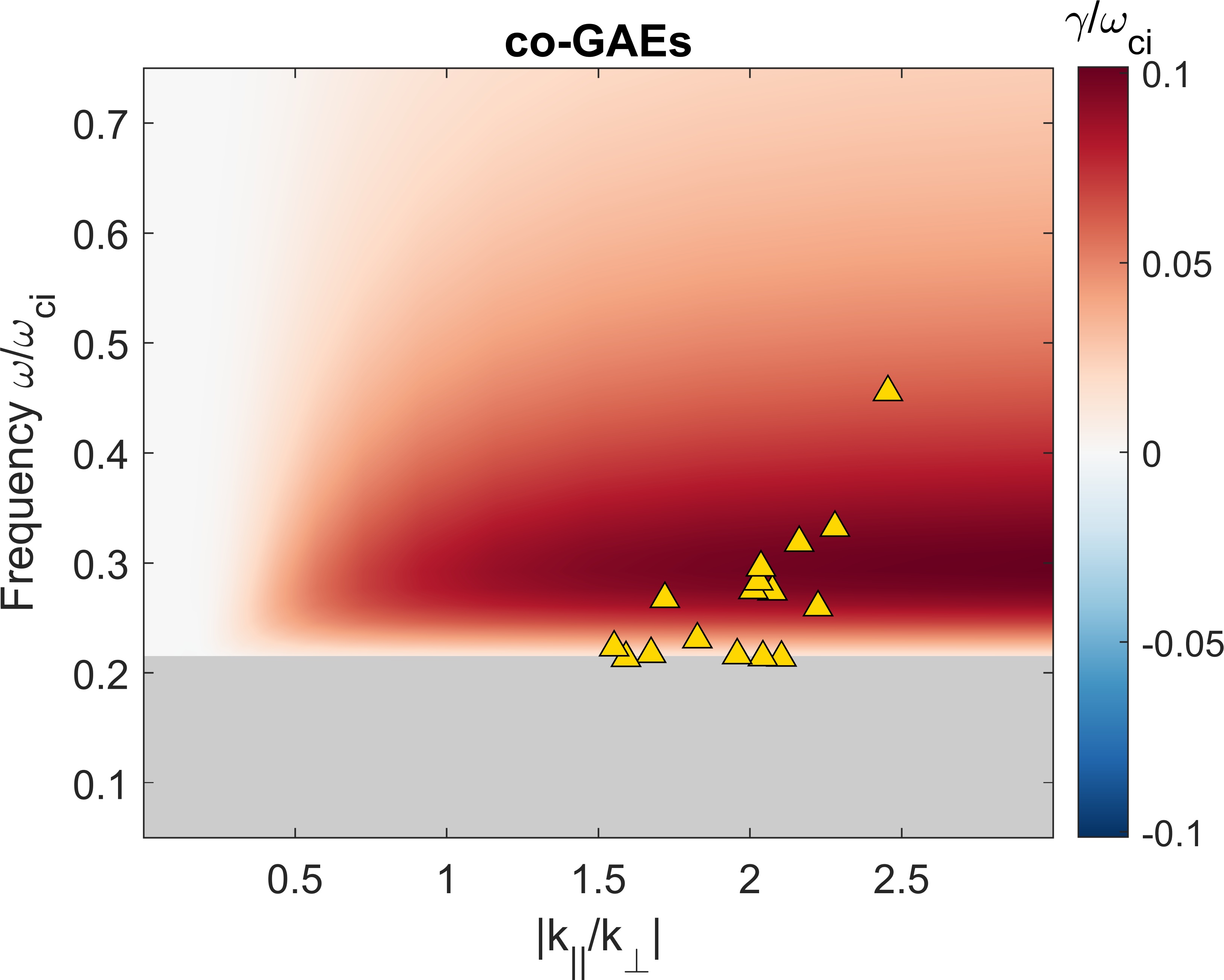}}
\caption{Analytically calculated growth rate (background color) as a function of frequency $(\omeganorm)$ and wave vector direction $(\krat)$ compared to the properties of unstable modes excited in simulations (gold markers). (a) co-CAEs driven by a fast ion distribution parameterized by $\linj = 0.7$ and $4.5 \leq \vinj \leq 5.5$, (b) cntr-GAEs driven by $\linj = 0.7$ and $3.0 \leq \vinj \leq 4.0$, and (c) co-GAEs driven by $\linj = 0.1$ and $5.0 \leq \vinj \leq 6.0$.}
\label{fig:ao_comp}
\end{figure}

A comparison between these conditions and the simulation results can be used to simultaneously verify theory and also understand the simulation results. Such a comparison is shown in \figref{fig:quant}, where $\vpres = (\omega - \lres\avg{\omegaci})/\kpar$ is determined from the resonance condition with $\kpar$ calculated from the mode structure in each simulation as described at the beginning of \secref{sec:id}. Each point represents an unstable mode from an individual simulation, with the beam injection velocity used in the simulation plotted against the marginal value of $\vinj$ necessary for instability, as determined by the preceding equations. The shaded regions -- green for co-GAEs, red for cntr-GAEs, and blue for co-CAEs -- indicate the regions of instability predicted by the theoretical conditions. For the GAEs, there is quite good agreement between theory and simulations, with only a few of the co-GAEs appearing far from the predicted boundary and all of the cntr-GAEs falling in the predicted region. 

The agreement for CAEs is not as strong, though the majority of unstable modes are still in the theoretically predicted region. There are three reasons why this might be the case. First, recall that the local theory which led to the predicted boundaries neglected the $\pphi$ gradient contribution to the fast ion drive/damping. As will be discussed briefly in \secref{sec:pphi}, this term gives a contribution proportional to $n\partial\fb/\partial\pphi$. Hence it provides additional drive for co-CAEs and co-GAEs $(n > 0)$, which would further extend the region of instability. Second, the tail of the distribution for $v > v_0$ was not accounted for in \eqref{eq:coCAE}. However, these particles can provide additional drive for co-propagating modes because they provide more resonant particles with $\lambda > \linj$ (corresponding to $\partial\fb/\partial\lambda < 0$). Lastly, the approximate determination of $\vpres \approx \omega/\kpar$ may be inadequate, since it assumes that sideband drift resonances are sub-dominant. Since much of the disagreement between theory and simulation occurs for beams with $\linj\gtrsim 0.5$, it may be that trapped particles are playing a more important role and that the sideband resonances need to be properly weighted, in addition to the primary resonance. 

Altogether, this comparison suggests that although the local theory is broadly sufficient for understanding the excitation of the different types of modes, some features of nonlocal theory are needed to improve the accuracy for the co-CAEs, which are generally closer to marginal stability where additional corrections could tip the scales.  

\subsection{Properties of unstable mode spectra} 
\label{sec:stabao}

Another approach for comparison between simulation and theory is to instead fix the beam parameters and consider how the growth rate depends on the properties of each mode, namely its frequency $\omeganorm$ and direction of wave vector $\krat$. Such analysis can be useful in the interpretation of experimental observations, since we will find that different types of modes are more unstable for different characteristic properties. This comparison is made in \figref{fig:ao_comp}. In these figures, the background color is the analytically computed growth rate, with darker red indicating larger predicted growth rate, and blue indicating regions with negative growth rate (damped by fast ions). Gray regions indicate where no resonance is possible, occurring when $\vb < \vpres$. The gold points represent unstable modes in \HYM simulations for beam distributions with fixed values of $\linj$ and $\vinj$ as specified on the plots for each type of mode. A small range of $\vinj$ around a central value is included in each case in order to include enough examples from the simulations to show a trend. 

For co-CAEs, a minimum value of $\krat$ is needed in order to resonate with the mode at all, and an even larger value is needed in order to have net fast ion drive. This is reasonable since combination of the $\lres = 0$ resonance condition and approximate CAE dispersion $\omega \approx k\va$ yields $\vpres/\va = \sqrt{1 + \kperp^2/\kpar^2}$. For a resonance to exist, $\vb > \vpres$ must be satisfied, which requires sufficiently large $\krat$ since this corresponds to sufficiently small $\vpres$. Meanwhile, the approximate instability condition written in \eqref{eq:coCAE} is of the form $\vb > \vpres g_c(\linj,\dl)$ where $0 < g_c(\linj,\dl) < 1$ depends on the beam injection geometry and weakly on the degree of anisotropy. Hence an even larger value of $\krat$ is necessary for the modes to be unstable than simply for the resonance to be satisfied. A maximum value of $\krat$ for the existence of co-CAEs can be derived heuristically based on the condition that the CAE is trapped within a magnetic well on the low field side \cite{Lestz2020p2}, which gives $\krat \lesssim n / 2\pi$, shown as a dashed line on \figref{fig:ao_comp_caez}. The agreement between simulations and theory in the figure is close but imperfect. While the bulk of the unstable modes from simulations cluster around the region of maximum growth rate, some of the modes are very close to or even on the wrong side of the predicted boundary. A likely explanation is the lack of $\pphi$ gradient in the theory calculation. 

The same comparison is very successful for both cntr- and co-GAEs. The figures illustrate a key difference between the unstable spectrum of GAEs vs CAEs. For co-CAEs, resonance requires sufficiently large value of $\krat$, with an even larger value needed for net fast ion drive. In contrast, the GAEs require sufficiently large \emph{frequency} to satisfy the resonance condition. Again, this can be understood as a consequence of the distinct dispersion relations for CAEs vs GAEs. For GAEs, $\omega \approx \abs{\kpar}\va$, so $\vpres/\va \approx \abs{\lres(\omegaci/\omega) - 1}$. Consequently, a resonance is possible when $\vb > \vpres$, which requires sufficiently large $\omeganorm$. For cntr-GAEs, the approximate condition for instability given in \eqref{eq:cntrGAE} takes the form $\vb < \vpres g_g(\linj)$, where $0 < g_g(\linj) < 1$. Consequently, excessively large frequencies can violate this condition (since they correspond to small $\vpres$), so theory predicts a band of unstable cntr-GAE frequencies. Nearly all of the simulation points fall within this band. 

In the case of co-GAEs, there is no maximum frequency for unstable modes since the inequality in their approximate instability condition is flipped relative to that for the cntr-GAEs. Hence, both the resonance condition and the instability condition provide upper bounds on $\vpres$, or equivalently lower bounds on $\omeganorm$. For the beam parameters shown in \figref{fig:ao_comp_gaep}, the lower bound on $\omeganorm$ coming from the instability condition is less restrictive than that coming from satisfaction of the resonance condition, so there are no frequencies where a resonance is possible but the mode is stable. However this is not always the case. For smaller values of $\vinj$ (for example $\vinj = 4.0$), such a band of stable frequencies exist that are damped by the fast ions. Consistency is found between the properties of unstable co-GAEs in simulations and those predicted by analytic theory, as all of the simulation points appear above the minimum frequency required for resonance, and with values of the wave vector direction $\krat$ near the region of largest growth rate. 

To summarize, analytic theory predicts that instability for co-CAEs occurs for modes with $\krat$ above a threshold value which depends on beam parameters. A maximum value of $\krat$ for CAEs can also be determined heuristically. Meanwhile, co-GAEs are unstable for frequencies that are sufficiently large. Lastly, the most unstable GAEs are predicted to occur within a specific range of frequencies. The unstable modes in \HYM simulations exhibit these properties in the majority of cases, with outliers likely explained by known limitations of the local analytic theory used to calculate the stability boundaries. 

\section{Growth Rate Dependence on Critical Velocity and Beam Anisotropy} 
\label{sec:vcdl}

Whereas the majority of the simulation parameter scan performed for this study focused on varying the beam injection geometry and velocity, there are other parameters in the fast ion distribution function that can be understood even with a less comprehensive set of simulations. Namely, the ratio of the critical velocity to the beam injection velocity $\vcrit$ controls the steepness of the distribution with respect to velocity, while the inverse width of the beam in velocity space $\dlinv$ controls the level of anisotropy. 

\begin{figure}[tb]
\includegraphics[width = \thirdwidth]{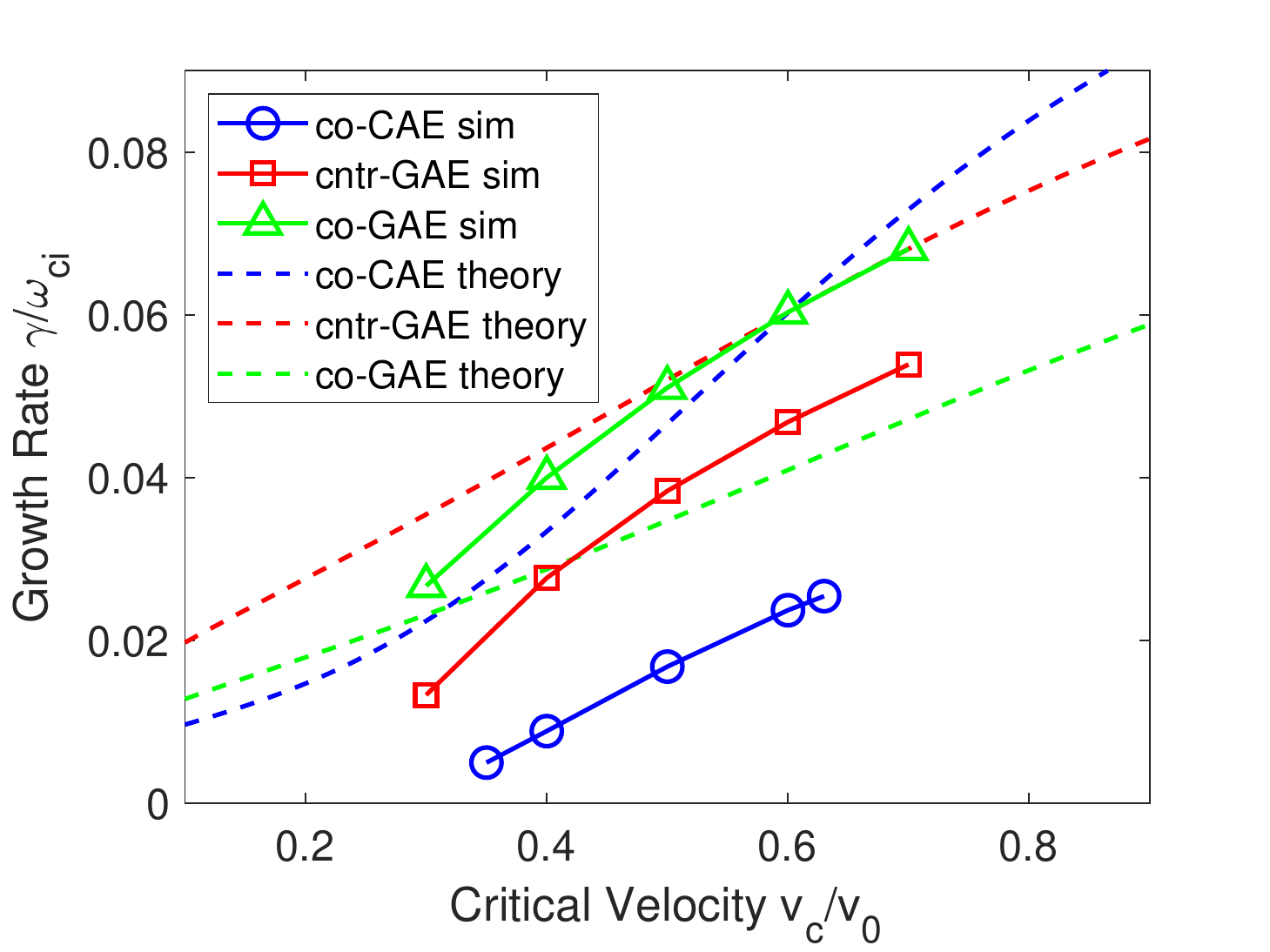}
\caption{Growth rate as a function of the normalized critical velocity $\vcrit$ from simulations (points connected by solid curves) and calculated by analytic theory (dashed curves). Blue curves are for co-CAEs, with $\linj = 0.7$, $\vinj = 5.5$, and restricted to $n = 4$. Red curves are for cntr-GAEs, with $\linj = 0.7$, $\vinj = 5.0$, and restricted to $n = -6$. Green curves are for co-GAEs, with $\linj = 0.1$, $\vinj = 5.0$, and restricted to $n = 9$. All simulations used $\dl = 0.3$.}
\label{fig:vc_scan}
\end{figure}

The dependence of the growth rate on $\vcrit$ is shown in \figref{fig:vc_scan} as determined by simulations and also calculated by the analytic expression previously discussed. It is clear that larger values of $\vcrit$ tend to make all modes more unstable -- whether they are CAEs or GAEs and co- or cntr-propagating. Since $\vc \propto \sqrt{\Te}$ \cite{Gaffey1976JPP}, this trend implies that a lower plasma temperature at fixed beam voltage reduces the fast ion drive. There are two key factors leading to this effect. First, increasing $\vcrit$ while keeping the total number of fast ions fixed  leads to a larger number of high energy resonant particles relative to those with low energy, thus providing more energy to drive the mode. Second, larger $\vcrit$ corresponds to smaller magnitude of $\partial\fb/\partial v$, so the fast ion ``damping'' from this term is also decreased. In \figref{fig:vc_scan}, the range of simulated values of $\vcrit$ is constrained by the self-consistent equilibrium solver and chosen background profiles. Note that the slopes of the analytic curves are quite similar to those of the simulation results, indicating similar dependence on $\vcrit$, even without quantitative agreement for all of the reasons previously discussed. 

However, the dependence of the growth rate on temperature may be more subtle in reality. For instance, prior DIII-D experiments at high field (1.9 T) found that sub-cyclotron modes were more easily excited at lower temperatures, likely due to the fast ion distribution becoming more isotropic at higher temperatures \cite{Heidbrink2006NF}. Moreover, the beam deposition profile could also be affected by the plasma temperature. Lastly, the mode damping on thermal electrons will also depend on their temperature, as discussed in \secref{sec:damp}, but this interaction is not captured in the simulation model used for this work. Hence, the trend of stronger drive with larger $\vcrit$ may be one of several factors which influence how CAE/GAE activity depends on background temperature. 

\begin{figure}[tb]
\includegraphics[width = \thirdwidth]{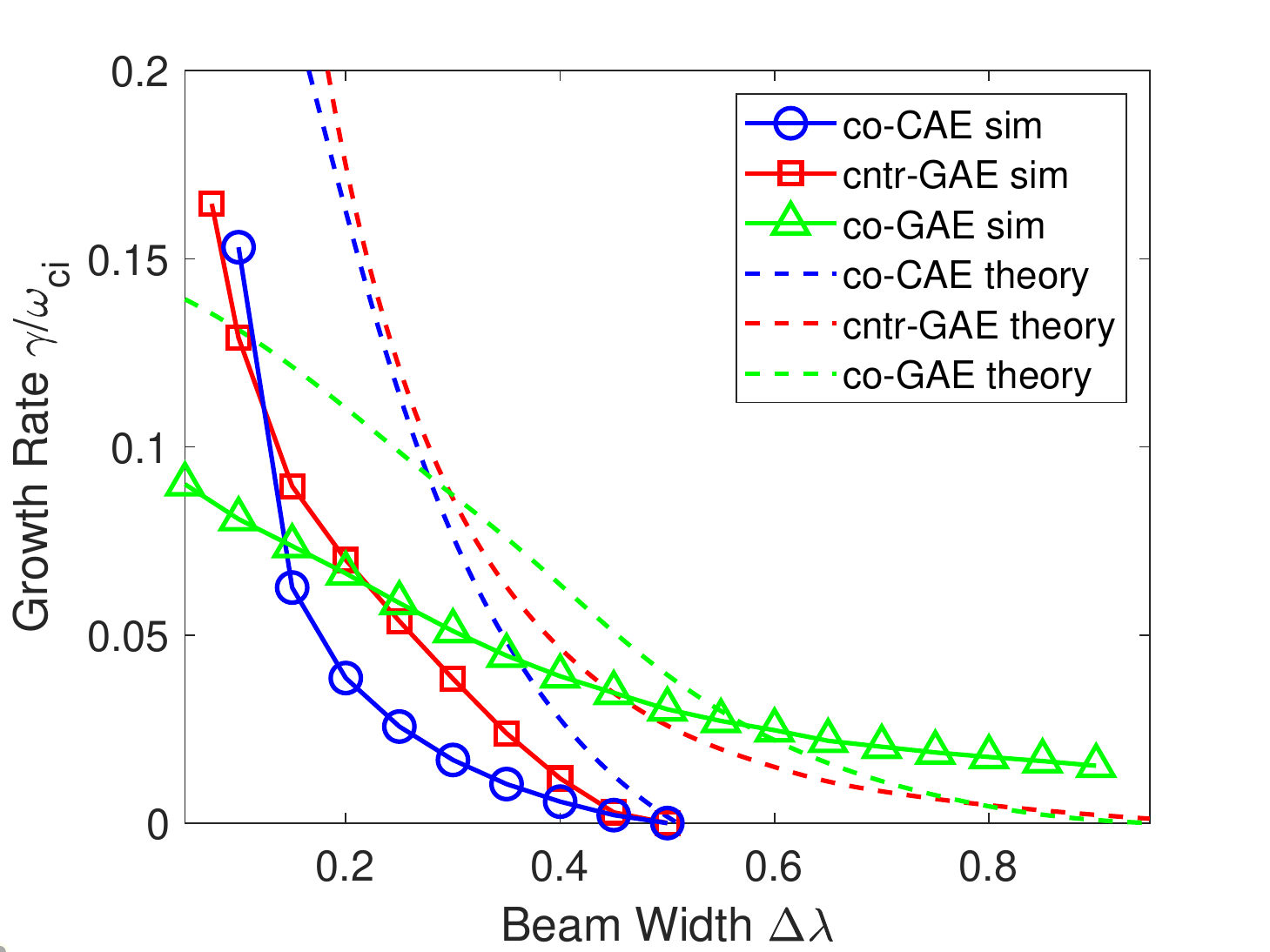}
\caption{Growth rate as a function of the beam width in velocity space $\dl$ from simulations (points connected by solid lines) and calculated by analytic theory (dashed lines). Blue curves are for co-CAEs, with $\linj = 0.7$, $\vinj = 5.5$, $\vcrit = 0.5$, and restricted to $n = 4$. Red curves are for cntr-GAEs, with $\linj = 0.7$, $\vinj = 5.0$, $\vcrit = 0.5$ and restricted to $n = -6$. Green curves are for co-GAEs, with $\linj = 0.1$, $\vinj = 5.0$, $\vcrit = 0.5$, and restricted to $n = 9$.}
\label{fig:dl_scan}
\end{figure}

Consider now the dependence on $\dlinv$, as shown in \figref{fig:dl_scan}. The simulations and analytic calculations agree that all types of modes become much more unstable as $\dl$ is decreased, consistent with the theoretical understanding that the dominant source of drive is beam anisotropy. Again, we find that there is qualitative but not quantitative agreement between the analytically calculated growth rates as a function of $\dl$ and those determined directly from simulations. To make the comparison shown in \figref{fig:dl_scan}, the normalized frequency $\omeganorm$ and wave vector direction $\krat$ were calculated from simulation results, and then a small range around these values was used to find the maximum analytic growth rate as $\dl$ was varied. 

Theoretically, the scaling with $\dl$ can be understood in two different regimes: the experimental regime where $\dl$ is relatively large $(\dl \gtrsim 0.2)$, and then the limiting case of $\dl \ll 1$. In \appref{app:dl}, it is demonstrated that $\gamma \propto 1/\dl^2$ when $\dl \gtrsim 0.2$, and $\gamma \propto 1/\dl$ in the limit of $\dl \rightarrow 0$. 

The scaling of the simulation results is approximately $1/\dl$ for the cntr-GAEs and co-GAEs, with a $1/\dl^2$ trend found for the co-CAEs. Interpretation of the simulation results is complicated by the presence of nontrivial damping in the simulations, which becomes more relevant as $\dl$ is increased. Hence it is consistent that the simulations would mostly capture the $\dl \ll 1$ scaling and have additional complications for the $\dl \gtrsim 0.2$ regime. It's unclear why the co-CAEs from the simulations exhibit the stronger $1/\dl^2$ dependence overall. 

It's also important to keep in mind that the analytic theory that we are comparing with is \emph{perturbative} in the sense that it assumes $\gamma \ll \omega$. Hence, any calculations predicting $\gamma \like \omega$ are explicitly unreliable. This places a lower bound on the value of $\dl$ that can be used to calculate the growth rate, as even $\dl \approx 0.2$ is giving $\gamma/\omegaci \approx 0.2$, corresponding to $\gamma/\omega \approx 0.6$ for sub-cyclotron frequencies $\omeganorm \approx 0.3$. The simulation model has no such restriction, but are still constrained to sufficiently large values of $\dl$ in order to satisfy constraints on the equilibrium for the modeled NSTX discharge. Whereas the examined co-CAE and cntr-GAE are stabilized for $\dl = 0.5$, the co-GAE remains unstable even for the largest simulated value of $\dl = 0.9$, which corresponds to extremely weak anisotropy. To explain this result, the previously neglected contribution from gradients with respect to $\pphi$ must be considered, as discussed in \secref{sec:pphi}.

\section{Corrections Due to Plasma Non-Uniformities \texorpdfstring{$(\partial\fb/\partial\pphi)$}{}}
\label{sec:pphi}

The analysis to this point has focused on fast ion drive resulting from fast ion velocity space anistropy or gradients in energy since these are the terms present in local analytic theory. However, a more complete treatment would also include the effects of plasma non-uniformities -- most notably a contribution from gradients with respect to $\pphi$. In this section, we will discuss the qualitative effect of this term that is absent in our theoretical analysis, and discuss its role in resolving certain disagreements between the self consistent simulation results and predictions from local analytic theory. 

A general form of the growth rate is adapted from \citeref{Kaufman1972PF} in \appref{app:pphi}. The relevant result is that 

\begin{align}
\gamma &\propto \int d\Gamma \left[\left(\frac{\lres \omegaci}{\omega} - \lambda\right)\pderiv{\fb}{\lambda} + \W\pderiv{\fb}{\W} + 
n\frac{\W}{\omega}\pderiv{\fb}{\pphi}\right]
\label{eq:gammapphi}
\end{align}

Here, $d\Gamma$ is the differential volume of phase space. The first two terms in brackets are the gradients present in the local theory (see \eqref{eq:gamma_dfdxdv}), while the third term results from plasma non-uniformity. An important consequence of \eqref{eq:gammapphi} is that the contribution to the growth rate from the $\partial\fb/\partial\pphi$ term depends on the sign of $n$. Hence, for non-hollow fast ion distributions ($\partial\fb/\partial\pphi > 0$ everywhere), it has a destabilizing effect for co-propagating modes $(n > 0)$ and a stabilizing effect for cntr-propagating modes $(n < 0)$. This consequence has been previously noted in the literature, and used to explain transitions between co- and cntr-propagation of toroidal \Alfven eigenmodes (TAEs) in TFTR \cite{Wong1999PLA} and NSTX-U \cite{Podesta2018NF}. This correction is expected to be most relevant for large values of $\abs{n}$. 

\begin{figure}[tb]
\includegraphics[width = \thirdwidth]{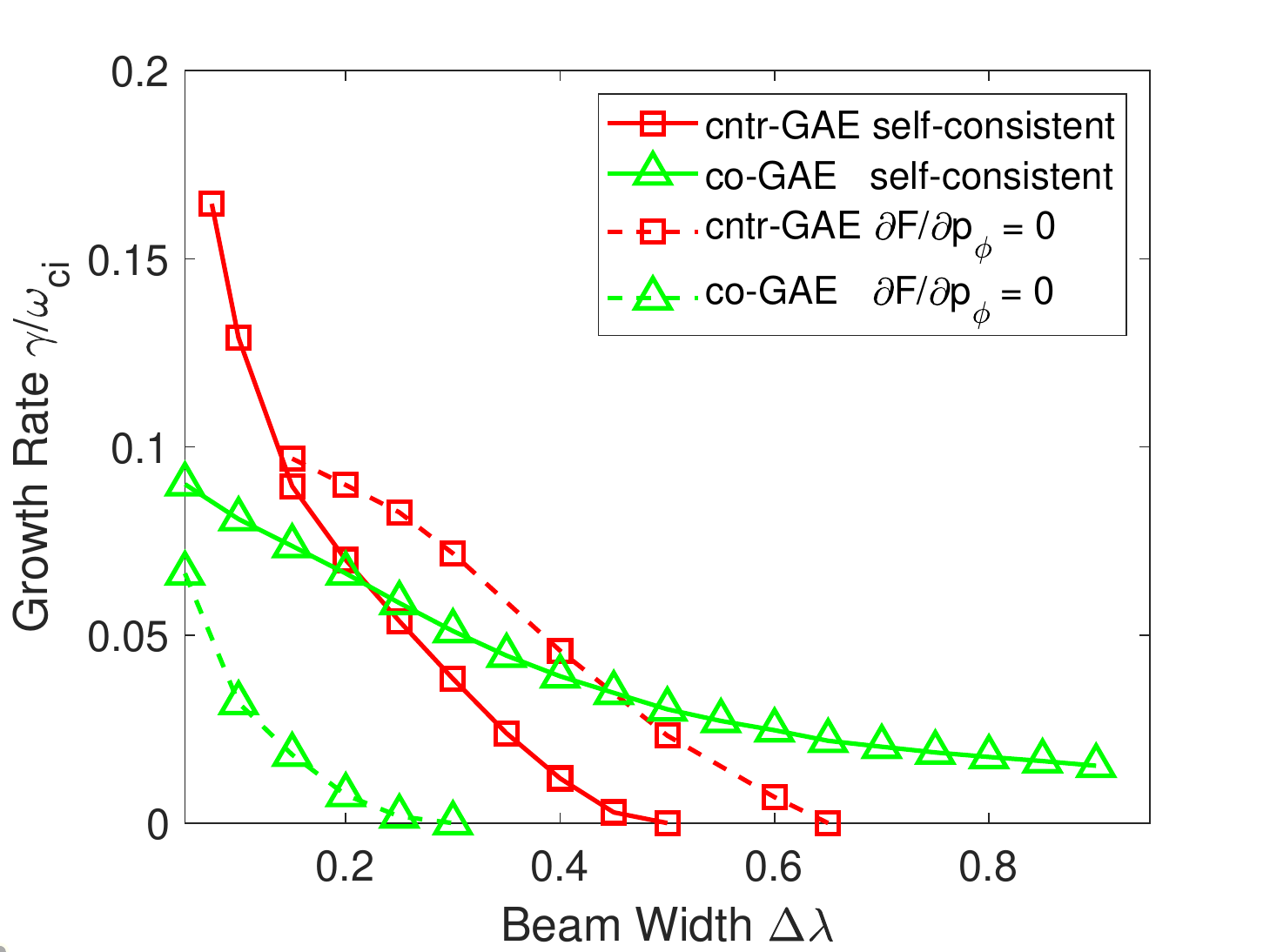}
\caption{Growth rate as a function of the beam width in velocity space $\dl$ from self-consistent simulations (solid lines, reproduced from \figref{fig:dl_scan}) and those excluding the effect of $\partial\fb/\partial\pphi$ (dashed lines). Red curves are for cntr-GAEs, with $\linj = 0.7$, $\vinj = 5.0$, $\vcrit = 0.5$ and restricted to $n = -6$. Green curves are for co-GAEs, with $\linj = 0.1$, $\vinj = 5.0$, $\vcrit = 0.5$, and restricted to $n = 9$.}
\label{fig:nopphi}
\end{figure}

To assess the importance of this correction, a set of non-self-consistent simulations was run where the $\pphi$ derivative was neglected from the time evolution equation for the particle weights (see \eqref{eq:dwdt}). In essence, this ``turns off'' the effect of $\partial\fb/\partial\pphi$ on the instability. Simulations were conducted with and without this term for cntr- and co-GAEs while varying $\dl$ in order to determine its influence on the growth rate, shown in \figref{fig:nopphi}. The solid curves are the same self-consistent simulation results as shown in \figref{fig:dl_scan}, while the dashed curves are ones where $\partial\fb/\partial\pphi = 0$ is imposed. It is immediately apparent that the gradient in $\pphi$ has a destabilizing effect for co-GAEs and stabilizing effect for cntr-GAEs, just as predicted by \eqref{eq:gammapphi}. Moreover, removing the contribution from $\partial\fb/\partial\pphi$ leads the co-GAE to be stabilized for $\dl > 0.3$, whereas its destabilizing effect supports the instability in self consistent simulations even when $\dl = 0.9$. It is reasonable that this effect might be strong for the co-GAEs which typically have large $\abs{n}$. This result suggests that co-GAEs may be driven unstable even by isotropic slowing down distributions (such as expected for energetic fusion products confined to the core), so long as the fast ions are sufficiently super-\Alfvenic ($\vinj \gtrsim 4$ in the simulated configuration) to satisfy the resonance condition for these modes.

When the same type of parameter scan was conducted for the co-CAEs shown in \figref{fig:dl_scan}, the modes were either stable or replaced by a more unstable cntr-GAE. This demonstrates that the gradient in $\pphi$ is essential to understanding conditions where the co-CAE is the most unstable modes -- it further destabilizes co-CAEs and damps cntr-GAEs that might otherwise have larger growth rate. This additional drive for co-CAEs from $\pphi$ which was neglected in the local analytic theory also explains why the approximate stability boundaries underestimated the fast ion drive for some co-CAEs, as seen in \figref{fig:quant_cae}. A similar but less significant disagreement was found for co-GAEs, as shown on \figref{fig:quant_gae}, which can be understood in the same way. However, the co-GAE's larger growth rates in general make this correction less likely to make the difference between stability and instability. 

\section{Rate of Mode Damping on the Thermal Plasma}
\label{sec:damp}

\begin{figure}[tb]
\includegraphics[width = \thirdwidth]{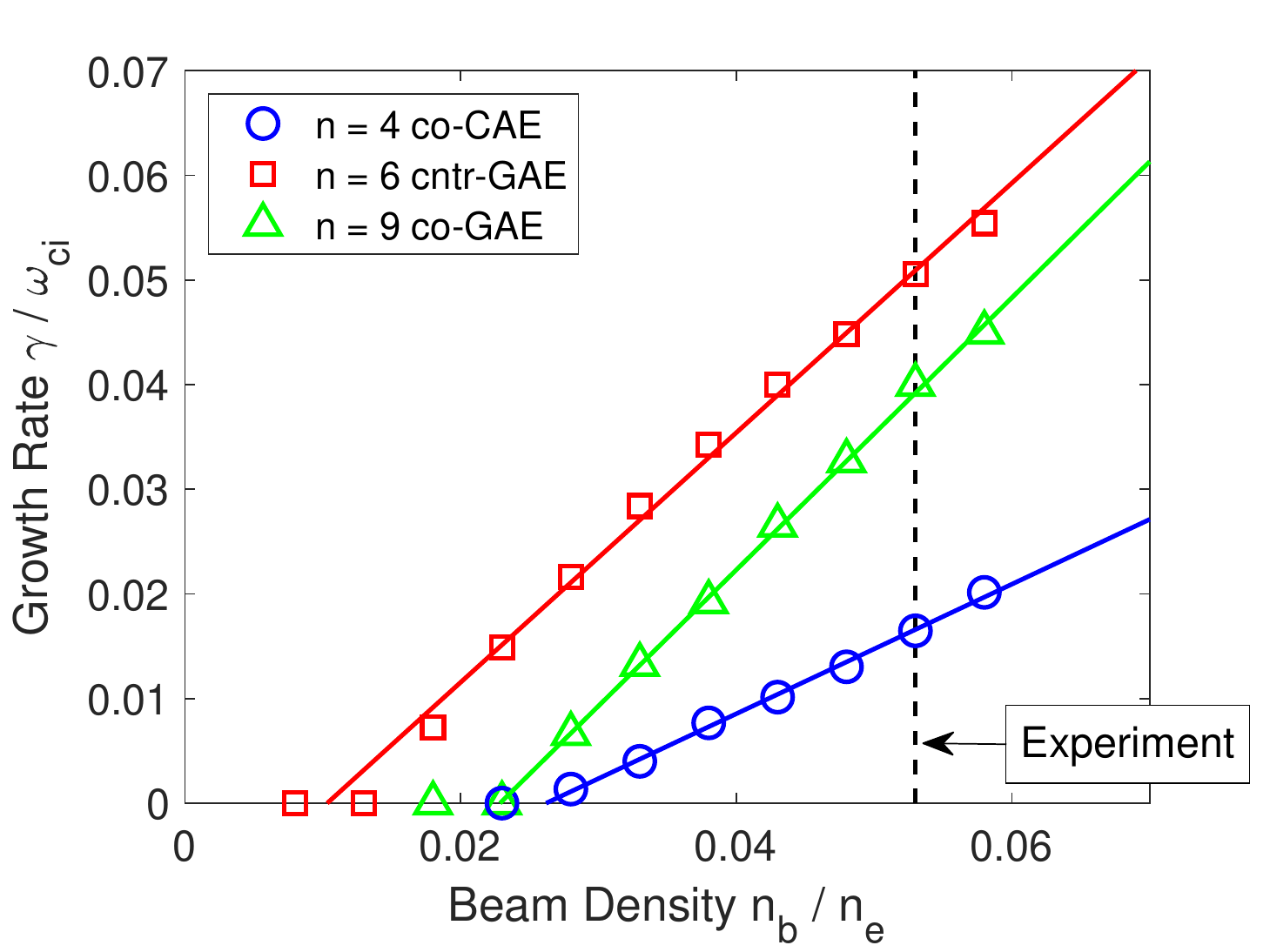}
\caption{Linear growth rates of representative cases of CAE/GAE modes as a function of beam density. Points are growth rates measured from simulations, while the lines are linear fits. Growth rate of zero indicates a simulation with no unstable mode. All simulations used $\vinj = 5.5$. Blue circles are an $n = 4$ co-CAE driven by $\linj = 0.7$, red squares are an $n = -6$ cntr-GAE driven by $\linj = 0.7$, and green triangles are an $n = 9$ co-GAE with $\linj = 0.3$ fast ions.}
\label{fig:growth_damping}
\end{figure}

In addition to the drive/damping that comes from the fast ions, the waves can also be damped due to interactions with the thermal plasma. Since the thermal plasma is modeled as a fluid, the simulation will necessarily lack damping due to kinetic effects, such as Landau damping and corrections to continuum damping due to kinetic thermal ions. The total damping rate present in the simulation for a specific mode can be determined by varying the beam density fraction. This is shown in \figref{fig:growth_damping} for one example of a co-CAE, cntr-GAE, and co-GAE. For each mode, there is a critical beam ion density $n_c/n_e$, below which the mode is stable. Above the critical density, the growth rate is proportional to density, as is expected in the perturbative regime where $\abs{\gamma} \ll \omega$. Hence, the relationship $\gamma_\text{net} = \gamma_0(n_b - n_c)/n_e$ is implied, allowing the inference of the damping rate $\gamma_0 n_c/n_e$. These critical densities imply thermal damping rates of $\gammadamp/\omegaci = 0.02 - 0.05$, corresponding to $20 - 60\%$ of the fast ion drive for the case with the nominal experimental fast ion density of $n_b/n_e = 0.053$. A beam density threshold for a cntr-GAE was recently demonstrated in DIII-D experiments, taking advantage of the unique variable beam perveance capability \cite{Tang2021U,Pace2018POP}.

Given this relatively large damping rate, it is natural to consider its primary source. The resistivity and viscosity in the simulations were varied to determine their influence on the damping. It was found that the damping rate was not very sensitive to either of these quantities. Changing the viscosity by an order of magnitude changed the total growth rate by a few percent, and changing the resistivity had an even smaller effect. Numerical damping could also be present in the simulations, though previous convergence studies of the growth rate for CAEs rules this out as a major source of damping. 

For CAEs, interaction with the \Alfven continuum has been previously identified as the likely dominant damping source, since mode conversion to a kinetic \Alfven wave near the \Alfven resonance location is apparent in the simulations \cite{Belova2015PRL,Belova2017POP}. For the GAEs, the robustness of the damping rate to the viscosity and resistivity also suggests that the primary damping source may be continuum damping. However, unlike the CAEs, coupling to the continuum is not always obvious in the mode structure. As investigated previously \cite{Lestz2018POP}, this may be due to the intrinsic non-perturbative nature of the GAEs -- they may fundamentally be energetic particle modes excited in the continuum, in which case their interaction with the continuum would be near the center of the mode instead of at its periphery, consequently obscuring the interaction. A more definitive identification of the GAE damping as due to its interaction with the continuum would require the calculation of the \Alfven continuum including the kinetic effects of thermal and fast ions, which is an avenue for further theoretical development. 

It is worthwhile to estimate the magnitude of the absent kinetic thermal damping and compare it to the sources present in the simulations. The thermal ions can be neglected because only a very small sub-population will have sufficient energy to resonate with the mode. However, due to the large ion to electron mass ratio, a large number of thermal electrons can interact with the mode. The total electron damping rate in a uniform plasma has been derived in \appref{app:damping}, assuming $\omega\ll\abs{\omegace},\omegape$, generalizing a derivation published by Stix in \citeref{Stix1975NF} which was restricted to $\kpar\ll\kperp$. In contrast, the modes in the simulations have $\krat = 0 - 3$, violating that assumption. 

\begin{figure}[tb]
\includegraphics[width = \halfwidth]{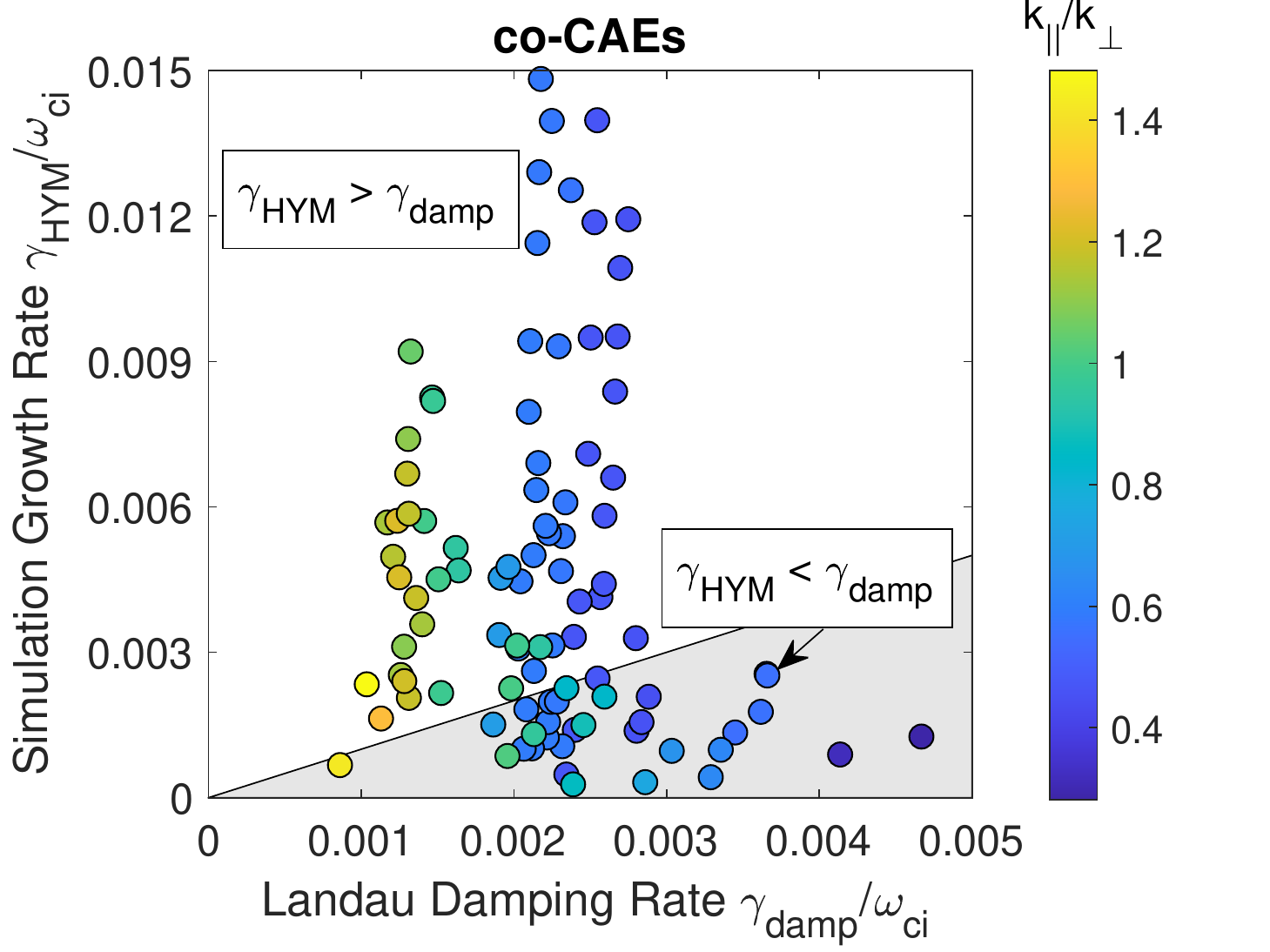}
\caption{Comparison of co-CAE growth rates from \HYM simulations and analytically calculated electron Landau damping rates. Shaded region indicates where electron damping (absent in simulation) would stabilize the mode. Color shows the value of $\krat$ for each mode, calculated from the mode structure in each simulation.}
\label{fig:landau_comp}
\end{figure} 

The general damping rate is given in \eqref{eq:dampfull}, including Landau damping, transit-time magnetic pumping, and their cross term. In order to recover the standard fast wave Landau damping rate, assume $\kpar\ll\kperp$, simplifying the result to 

\begin{align}
\lim_{\kpar \ll \kperp} \frac{\gammadamp^{\text{CAE}}}{\omega} = -\frac{\betae\sqrt{\pi} y e^{-y^2}}{2}
\end{align}

Here, $y = \omega/\kpar\vtherme$ and $\betae = 2\mu_0 P_e / B^2$. This is the familiar formula from \citeref{Stix1975NF}. \eqref{eq:dampfull} can also be used to derive intuition in the opposing limit of $\kpar \gg \kperp$. Then, defining $\omegabar \defined \omega/\omegaci$, one finds

\begin{align}
\lim_{\kpar \gg \kperp} \frac{\gammadamp}{\omega} = -\frac{\betae\sqrt{\pi} y e^{-y^2}}{2}\frac{\kperp^2}{\kpar^2}\frac{1 \pm 2\omegabar + 2\omegabar^2}{(2\pm\omegabar)(1\pm\omegabar)}
\end{align}

Above, the ``$+$'' corresponds to the compressional wave damping rate and the ``$-$'' is for shear waves. Hence electron damping scales like $\gammadamp \propto \kperp^2/\kpar^2$, favoring modes with larger values of $\krat$. The general CAE damping rate is mostly sensitive to $\krat$, depending very weakly on frequency. The maximum CAE damping rate occurs with a sharp peak at $y = 1/\sqrt{2}$, corresponding to $\alphacrithym = \sqrt{2 m_e / m_i \betae} \ll 1$, hence the maximum damping rate is $\gammadamp^{\text{CAE}}/\omega \leq \sqrt{\pi/8e}\betae = 0.38\betae$. However, most CAEs from the simulations have $\krat\gg\alphacrithym$, reducing the expected electron damping rate. For shear modes, the maximum growth rate typically occurs at some $\krat \like \ord{1}$. Unlike the compressional modes, the damping rate depends strongly on both $\krat$ and $\omeganorm$. Numerical evaluation of the analytic expression shows that $\gammadamp^{\text{GAE}}/\omega \leq 0.002$ for all values of $\betae < 1$. 

To estimate the importance of the electron damping for the modes studied in the \HYM simulations, we can evaluate the electron damping rate expression in \eqref{eq:dampfull} numerically without any approximation, using $\omeganorm$ and $\krat$ for each mode from the simulation, and $\betae = 8\%$ on-axis for each. For the GAEs, this exercise shows that the absent electron damping is relatively insignificant -- at most $10\%$ of the net growth rate in the simulation, and in most cases less than $0.1 - 1\%$. In contrast, the unidentified source of damping present in the simulation (likely continuum) is approximately $50\%$ of the net growth rate, so the hybrid model in \HYM which ignores kinetic electron effects is well-justified for calculations of GAE stability. 

For the CAEs, the electron damping can be more important, as shown in \figref{fig:landau_comp} where the net drive of each CAE in the simulation is compared to the analytically calculated electron damping rate. The shaded region shows where the electron damping exceeds the net drive in the simulations. This indicates that in some cases the electron damping missing from the hybrid model could be large enough to stabilize some of the marginally unstable CAEs in the simulations. On the other hand, for each distinct compressional eigenmode (unique frequency and mode structure) excited in this set of simulations, there exists a set of fast ion parameters sufficient to overcome the neglected electron damping. So while the quantitative growth rates should be corrected for the missing electron contribution, this neglected source of damping will not qualitatively change the instability picture. 

While this section considered the collisionless electron damping in a uniform plasma, further improvements could be made in future work by considering the effects of trapped particles, non-uniform geometry, and collisions. As discussed in \citeref{Gorelenkov2002NF}, the inclusion of trapped particles tends to reduce the overall damping rate. Work has also been done by Grishanov and co-authors \cite{Grishanov2001PPCF,Grishanov2002POP,Grishanov2003PPCF} to calculate the electron Landau damping in realistic low aspect ratio geometry, though further work must be done to include transit-time magnetic pumping and cross terms. Lastly, the effect of collisional trapped electron damping has previously been considered for TAEs, which typically enhances the damping rate \cite{Gorelenkov1992PS}. However, these results can not presently be applied outright to GAEs due to the different frequency range, nor to CAEs due to differences in dispersion and polarization. 

\section{Summary and Discussion}
\label{sec:summary}

Linear hybrid initial value simulations of NSTX-like, low aspect ratio plasmas were performed in order to investigate the influence of fast ion parameters on the stability and spectral properties of sub-cyclotron compressional (CAE) and global (GAE) \Alfven eigenmodes driven by the general Doppler-shifted cyclotron resonance condition. The model NBI distribution included several parameters that were studied: injection geometry $\linj$, normalized injection velocity $\vinj$, normalized critical velocity $\vcrit$, and the degree of velocity space anisotropy $\dlinv$.  

The simulations demonstrated unstable co-propagating CAEs, co-propagating GAEs, and cntr-propagating GAEs across many toroidal harmonics $\abs{n} = 3 - 12$ in the broad frequency range $\omeganorm = 0.05 - 0.70$ with normalized growth rates of $\gamma/\omegaci = 10^{-4} - 10^{-2}$. While both co- and cntr-propagating CAEs have been identified in NSTX experiments \cite{Fredrickson2013POP}, the cntr-CAEs never appear as the most unstable mode in \HYM simulations due to the initial value nature of the simulations. GAEs in the simulation were present in both co- and counter-propagating varieties. The cntr-GAEs have similar frequencies and toroidal harmonics as those observed in the model discharge for these simulations \cite{Belova2017POP}. Meanwhile, the co-GAEs are higher frequency and have not previously been observed in NSTX, likely due to beam geometry constraints. The co-GAEs should be excitable with the new off-axis beam sources installed on NSTX-U, given a discharge with sufficiently large $\vinj$. Fascinatingly, the GAEs in simulations behaved more similarly to energetic particle modes than perturbative MHD modes, in contrast to the simulated CAEs which were qualitatively consistent with eigenmodes found by the spectral Hall MHD code \CAETB. Moreover, it was found that a large fraction of unstable modes violated the common large aspect ratio assumption of $\kpar \ll \kperp$, instead having $\krat = 0.5 - 3$. These properties should be taken into account when interpreting experimental measurements where diagnostic limitations require the use of well-informed assumptions. 

The simulations revealed that cntr-propagating GAEs can be excited at a significantly lower $\vinj$ than the co-propagating CAEs and GAEs due to the different resonances which govern their excitation. In terms of beam geometry, it was found that the cntr-GAEs prefer perpendicular injection $(\linj \rightarrow 1)$, the co-GAEs prefer tangential injection $(\linj \rightarrow 0)$, and co-CAEs are most unstable for a moderate value of $\linj \approx 0.5$. Combination of NSTX-U's new tangential beam source with its original more radial one should provide sufficient flexibility to test this growth rate dependency. In addition, the GAEs typically have larger growth rates than the co-CAEs by about an order of magnitude. Due to the increased nominal on-axis magnetic field on NSTX-U, this result indicates that typical NSTX-U conditions should favor unstable GAEs over CAEs even more heavily than in NSTX. Early NSTX-U discharges appear to corroborate this conclusion \cite{Fredrickson2018NF}, though confirmation awaits more extensive operations. A local analytic theory \cite{Lestz2020p1,Lestz2020p2} for fast ion drive was used to interpret the growth rate trends. For typical NSTX NBI parameters, the gradient due to velocity anisotropy dominates, and clearly explains why the different types of modes become most unstable for different beam geometries. Similarly, the smaller growth rates for co-CAEs can be explained by the large factor $\omegaci/\omega$ which multiplies the growth rate of the co- and cntr-GAEs driven by the $\lres = \pm 1$ resonance. The theory also predicts that the growth rate of the most unstable mode should increase with $\vinj$, similar to the simulation findings. 

Good agreement between simulation results and approximate instability conditions was found for the GAEs in terms of the beam parameters necessary to excite the modes. The agreement in this area was not as strong for co-CAEs, though this is consistent with a more general, non-local theory including gradients in the fast ion distribution due to $\pphi$. Inclusion of this term provides an additional source of drive for co-propagating modes (such as the CAEs, which are sometimes stable without it), while it damps those that cntr-propagate. These instability conditions also predict that unstable co-CAEs must have a value of $\krat$ exceeding a certain threshold, with a maximum value determined heuristically. The unstable co-CAEs in simulations neatly fall within or very near this range. In contrast, the GAE instability conditions imply a narrow range of unstable frequencies for cntr-GAEs and a minimum frequency for the co-GAEs. The spectrum of unstable GAEs in both simulations and experiments \cite{Lestz2020p1} can be quantitatively explained through this lens. 

The growth rate dependence on the normalized critical velocity $\vcrit$ and beam anistropy $\dlinv$ determined in simulations could also be explained by theory. For all modes, increasing the parameter $\vcrit$ led to larger growth rates in simulations, indicating larger fast ion drive at higher plasma temperatures in proportion to $\sqrt{T_e/\W_\text{beam}}$. Larger fast ion anisotropy in velocity space also further destabilized all modes in the  simulations, similar to the $\gamma \like \dl^{-2} \tto \dl^{-1}$ scaling predicted analytically. Interestingly, it is found that the large $n$ co-GAEs receive substantial drive from $\partial\fb/\partial\pphi$, allowing them to remain unstable when there is much weaker anisotropy than required to drive the cntr-GAEs and co-CAEs. 

Lastly, an assessment of the background damping was made. In simulations, background damping rates of $20 - 60\%$ of the net drive was found for the beam parameters most closely matching the modeled experimental conditions. This damping has been attributed to radiative and continuum damping in previous analysis \cite{Belova2015PRL,Belova2017POP}. For GAEs the source is less certain, but past theoretical work has concluded that continuum damping is likely the primary mechanism \cite{Gorelenkov2003NF}. The electron damping absent in the \HYM model has also been estimated analytically, indicating that it should be negligible relative to the fast ion drive of GAEs in all cases, but could be comparable for a subset of the more weakly driven co-CAEs in simulations.

Together, the large set of simulations combined with their broadly successful theoretical interpretation within a simple, transparent theoretical framework helps build intuition for how the spectrum of unstable CAEs and GAEs is influenced by the properties of the fast ion distribution. The  information gathered about the properties of the modes can also be leveraged to help guide the notoriously difficult task of distinguishing CAEs from GAEs in experimental observations \cite{Crocker2013NF}. In addition, the results presented here suggest some new avenues of inquiry for investigating the anomalously flat electron temperature profiles observed on NSTX at high beam power, which correlated with strong CAE/GAE activity. With the large beam parameter space on available on NSTX-U, these insights can help guide future experiments to perhaps isolate the effects of CAEs vs GAEs on the enhanced transport in order to distinguish between different transport mechanisms. 

A detailed simulation study of multi-beam effects on CAE/GAE stability is a compelling next step, especially when considering the very robust stabilization of GAEs found with the off-axis, tangential beam sources on NSTX-U \cite{Fredrickson2017PRL,Fredrickson2018NF,Belova2019POP}. While this paper focused on the influence of fast ion parameters, the stability properties also depend on the equilibrium profiles, which could be explored in future work. Development of a complete nonlocal analytic theory including both fast ion drive and relevant background damping sources would be the next step forward for the local theory used here which works well qualitatively but does not give reliable values for the total growth rate. Ideally, the combination of the present work with these additional steps could enable the development of a reduced model for predicting the most unstable CAE and GAE modes in a given discharge scenario, en route to a more complete understanding of their influence on the electron temperature profiles.

\section{Acknowledgements}
\label{sec:Acknowledgements}

JBL thanks V.N. Duarte for fruitful discussions. The simulations reported here were performed with computing resources at the National Energy Research Scientific Computing Center (NERSC). The data required to generate the figures in this paper are archived in the NSTX-U Data Repository ARK at the following address: \url{http://arks.princeton.edu/ark:/88435/dsp011v53k0334}. This research was supported by the U.S. Department of Energy (NSTX contract \# DE-AC02-09CH11466).

\appendix 

\section{Growth rate scaling with beam anisotropy}
\label{app:dl}

The purpose of this derivation is to determine the overall scaling of the growth rate with the beam anisotropy parameter $\dl$. To do so, we will consider the dominant contribution from anisotropy alone in \eqref{eq:gamma_dfdxdv} in the small FLR limit for $\lres = 1$ GAEs. This yields

\begin{align}
\gamma \appropto \frac{1}{C_f(\dl)\dl^2}\int_0^{\omegacires\lm} \frac{\lambda(\lambda-\linj)}{(1-\lambda\omegacires)^2}
e^{-(\lambda-\linj)^2/\dl^2} d\lambda
\end{align}

Here, the upper limit of integration $\lm = 1 - \vpres^2/\vb^2$ is approximated as $\lm \approx \linj$, approximately corresponding to the condition for largest growth rate. $C_f(\dl)$ is a normalization constant to keep the fast ion density constant. For large $\dl$ such that $\linj - \dl\sqrt{2} \gtrsim 0$, the Gaussian dependence on $\lambda$ in \eqref{eq:F2} is very weak and can be approximated by a constant, which also removes the $\dl$ dependence from the normalization constant $C_f$. Then we have 

\begin{align}
\gamma \appropto \frac{1}{\dl^2}\int_0^{\linj} \frac{\lambda(\lambda-\linj)}{(1-\lambda\omegacires)^2} d\lambda \like \frac{1}{\dl^2}
\end{align}

Conversely, consider very small $\dl$ where the distribution is so narrow that only the behavior of the integrand very close to $\lambda \approx \linj$ is relevant. Then the normalization with respect to $\dl$ can be approximated as

\begin{align}
C_f^{-1} = \int_0^{\omegacires^{-1}} \frac{e^{-(\lambda-\linj)^2/\dl^2}}{\sqrt{1-\lambda\omegacires}}d\lambda 
\approx \frac{\dl\sqrt{\pi}}{\sqrt{1 - \linj\omegacires}}
\end{align}

Subsequent Taylor expansion of the rest of the integrand gives $\lambda(\lambda-\linj)/(1 - \lambda\omegacires)^2 \approx \linj(\lambda-\linj)/(1 - \linj\omegacires)^2$ permits integration:

\begin{align}
\gamma &\appropto \frac{1}{\dl^3}\frac{\linj}{(1 - \linj\omegacires)^2}\int_0^{\linj} (\lambda - \linj)e^{-(\lambda-\linj)^2/\dl^2} d\lambda \\ 
&= \frac{1}{d\lambda}\frac{\linj}{2(1 - \linj\omegacires)^2}\left(-1 + e^{-\linj^2/\dl^2}\right) \like \frac{1}{\dl}
\end{align}

Numerical evaluation of the unapproximated analytic expression for fast ion drive confirms that the growth rate scales as $1/\dl^2$ for $\dl \gtrsim 0.2$, transitioning to a different asymptotic scaling of $1/\dl$ in the limit of $\dl \ll 1$. Analogous arguments to those given above can also be made for the $\lres = 0$ resonance (relevant for CAEs), which result in the same scalings. Similar arguments can be made for the large FLR limit, which does not affect the result. For concreteness, simulations show that typically $\omegacires \approx 0.9$. 

\section{Growth rate correction due to gradients in \texorpdfstring{$\pphi$}{pphi}}
\label{app:pphi}

A general form of the growth rate is given by Kaufman in Eq. 37 of \citeref{Kaufman1972PF} in terms of action angle coordinates as 

\begin{align}
\gamma &\propto \int d\Gamma \left(\vec{\lres}\dot\grad_{\vec{J}}\right) \fb
\label{eq:kaufstart}
\end{align}

Here, $d\Gamma$ is the differential volume of phase space, $\vec{\lres} = \left(\lres, n, \lres_P\right)$ is the vector of integers for the resonance condition, and $\vec{J} = \left(\mu, \pphi, J_P\right)$ is the vector of actions. $J_P$ is a constant motion defined as an integral over poloidal motion (see Eq. 13 of \citeref{Kaufman1972PF}). There are additional terms present in the integrand, but we will ignore those for now since the aim of this section is to obtain qualitative understanding of the effect of the gradient in $\pphi$. The chain rule can be used to transform \eqref{eq:kaufstart} into the variables $(\lambda,\pphi,\W)$: 

\begin{align}
\label{eq:kaufpphia}
\gamma &\appropto \int d\Gamma \left[ \lres \pderiv{\fb}{\mu} + n\pderiv{\fb}{\pphi} + \left(\lres\pderiv{\W}{\mu} + n\pderiv{\W}{\pphi} + \lres_P\pderiv{\W}{J_P}\right)\pderiv{\fb}{\W} \right] \\ 
\label{eq:kaufpphib}
&= \int d\Gamma \left[ \lres \pderiv{\fb}{\mu} + n\pderiv{\fb}{\pphi} + \omega\pderiv{\fb}{\W} \right] \\ 
&= \int d\Gamma  \frac{\omega}{\W}\left[\left(\frac{\lres\omegaci}{\omega} - \lambda\right)\pderiv{\fb}{\lambda} + \W\pderiv{\fb}{\W} + n\frac{\W}{\omega}\pderiv{\fb}{\pphi}\right]
\label{eq:kaufpphi}
\end{align}

An alternative (and equivalent) form of the resonance condition was used to simplify from \eqref{eq:kaufpphia} to \eqref{eq:kaufpphib}: $\omega = \lres(\partial\W/\partial\mu) + n(\partial\W/\partial\pphi) + \lres_P(\partial\W/\partial J_P)$, as in Eq. 30 in \citeref{Kaufman1972PF}. The form of \eqref{eq:kaufpphi} is consistent with a similar expression found in \citeref{Wong1999PLA}, \citeref{Coppi1986PF}, \citeref{Gorelenkov1995POP}, and \citeref{Kolesnichenko2006POP}, which use an approximation $m_i \vpar \ll e Z_i \psi'(r)$ in order to re-write $\partial\fb/\partial\pphi \approx - [q / (\omegaci m_i r)](\partial\fb/\partial r)$. Note that \citeref{Kolesnichenko2006POP} uses an opposite convention for the sign of $n$ from what is used in other works (including this one), leading to a relative sign difference. 

\section{General expression for electron damping of compressional and shear \Alfven eigenmodes}
\label{app:damping}

In this appendix, the electron damping rate for a uniform plasma is generalized from \citeref{Stix1975NF} to include the cases where $\kpar \ll \kperp$ is not satisfied. Consequently, this rate will include Landau damping, transit-time magnetic pumping, as well as their cross term. The normalized damping rate is given by $\gammadamp/\omega = -\Pwave/\omega \Wwave$, where $\Pwave$ is the power density transferred to the particles from the wave, and $\Wwave$ is the wave energy density. To ensure accuracy, the complete two-fluid dispersion instead of the approximate forms $\omega \approx k\va$ (CAEs) and $\omega \approx \kpar\va$ (GAEs) will be used. In a uniform geometry with $B_0$ oriented in the $\hat{z}$ direction, and $\kperp$ in the $\hat{x}$ direction, the cold plasma dispersion is determined by 

\begin{align}
\begin{pmatrix}
\Kxx - \npar^2 & \Kxy & \npar\nperp \\ 
\Kyx & \Kyy - n^2 & 0  \\ 
\npar\nperp & 0 & \Kzz - \nperp^2
\end{pmatrix}
\begin{pmatrix}
E_x \\ E_y \\ E_z 
\end{pmatrix}
= 0
\label{eq:wavetens}
\end{align}

Above, $n = k c / \omega$ is the index of refraction. The directions are defined such that $\vec{k} = \kperp\hat{x} + \kpar\hat{z}$. $\Kij$ are the usual cold plasma dielectric tensor elements \cite{StixWaves}. As explained by Stix, the low frequency, high conductivity limit gives the MHD condition $\Epar \ll \Eperp$. Again taking the low frequency limit ($\omega \ll \omegape,\abs{\omegace}$), defining $\omegabar \defined \omega/\omegaci$, we have 

\begin{subequations}
\begin{align}
\label{eq:tens-S}
\Kxx &= \Kyy = S \approx \frac{c^2}{\va^2}A \\ 
\label{eq:tens-D}
\Kxy &= -\Kyx = -i D \approx i\omegabar\frac{c^2}{\va^2}A \\ 
\label{eq:tens-A}
\text{where } A &\defined \frac{1}{1 - \omegabar^2}
\end{align}
\label{eq:wavetens-cold}
\end{subequations}

Define the \Alfven refractive index $N = k\va/\omega$, $F^2 = \kpar^2/k^2$, and $G = 1 + F^2$. Then the two-fluid coupled dispersion is readily found by neglecting $E_z$: 

\begin{align}
N^2 &= \frac{A G}{2 F^2}\left[1 \pm \sqrt{1 - \frac{4 F^2}{A G^2}}\right]
\label{eq:stixdisp}
\end{align}

For $\omega < \omegaci$, the ``$+$'' solution corresponds to the shear wave, and ``$-$'' solution to the compressional wave. For $\omega > \omegaci$, the shear wave does not propagate, and the ``$+$'' solution corresponds to the compressional wave. The polarization will be needed to compute $\Pwave$ and $\Wwave$. The second line of \eqref{eq:wavetens} gives 

\begin{equation}
H \defined i\frac{E_x}{E_y} = -\frac{1}{\omegabar}\left(\frac{N^2}{A} - 1\right)
\label{eq:Hexp}
\end{equation}

While $\Kzy$ and $\Kzz$ were not needed to calculate the cold dispersion, their hot forms are needed to accurately capture the $\Epar$ effects. In the limit of $\omega^2/\kpar^2\vtherme^2 \ll 1$, which is the regime studied here, the relevant hot tensor elements are 

\begin{align}
\Kzy &\approx \frac{i\kperp\omegape^2}{\kpar\omega\abs{\omegace}} = \frac{i}{\omegabar}\abs{\frac{\kperp}{\kpar}}\frac{c^2}{\va^2} \\ 
\Kzz &\approx \frac{1}{\kpar^2\ldebye^2} = \frac{2}{\Npar^2\omegabar^2\betae}\frac{c^2}{\va^2}
\end{align}

Note $\omegape^2 = n_e e^2 / \epsilon_0 m_e$ is the electron plasma frequency, $\ldebye^2 = \epsilon_0 T_e/n_e e^2$ defines the Debye Length, $\vtherme = \sqrt{2 T_e/m_e}$ defines the thermal electron velocity, and $\betae = 2 \mu_0 n_e T_e / B^2$ is the electron pressure normalized to the magnetic pressure. Note the following relations which are useful for simplifying expressions above and below: $\omegaci\vtherme^2 = \betae\abs{\omegace}\va^2$, $\omegape^2\va^2 = c^2\omegaci\abs{\omegace}$, and $2\ldebye\omegape = \vtherme^2$. The first and third lines of \eqref{eq:wavetens}, modified to include the hot forms of $\Kzy$ and $\Kzz$, implies 

\begin{align}
J &\defined \frac{E_z}{E_y} = \frac{-\npar\nperp \Kxy + (\Kxx - \npar^2)\Kzy}{\npar^2\nperp^2 - (\Kxx - \npar^2)(\Kzz - \npar^2)} \\ 
&= i\left[\frac{-\omegabar A \Nperp\Npar + (A - \Npar^2)\abs{\kperp}/\left(\abs{\kpar}\omegabar\right)}{\Nperp^2\Npar^2 - (A - \Npar^2)\left(\frac{2}{\beta\omegabar^2\Npar^2} - \Nperp^2\right)}\right]
\label{eq:Jexp}
\end{align}

The absorbed power density $\Pwave$ is given by 

\begin{align}
\Pwave &= -\frac{i\omega}{16\pi}\vE\conj\dot\Ksym\dot\vE + \text{ c.c.} \\
&= -\frac{i\omega}{8\pi}\Re{\vE\conj\dot\Ksym\antiherm\dot\vE} \\ 
&= -\frac{i\omega\abs{J}}{8\pi}\left[\Kyy\antiherm + 2 i \Kyz\antiherm\Im{J} + \Kzz\antiherm\abs{J}^2\right] 
\label{eq:pwave-raw}
\end{align}

Above, $\Ksym$ is the hot dielectric tensor and $\Ksym\antiherm$ is its anti-Hermitian part. We also define $\vE$ such that the real wave field $\vE_0 = \Re{\vE e^{i(\vk\dot\vec{x} - \omega t)}}$. Only the nonzero terms for the $\omega - \kpar\vpar = 0$ resonance are kept since $\omega\ll\abs{\omegace}$. Contributions from higher resonances will be smaller by a factor of $\exp(-\omegace^2/\kpar^2\vtherme^2) \ll 1$. The anti-Hermitian tensor elements are given in Eq. 12 in \citeref{Stix1975NF} and then simplified as 

\newcommand{\Gstix}{Q}
\begin{align}
\Kyy\antiherm &= \frac{i\kperp^2\vtherme^2}{\omega\abs{\omegace}}\Gstix
= i\omegabar\Nperp^2\betae \Gstix \\ 
\Kyz\antiherm &= \abs{\frac{\kperp}{\kpar}}\Gstix \\ 
\Kzz\antiherm &= \frac{2i\omega\abs{\omegace}}{\kpar^2\vtherme^2}\Gstix
= \frac{2i \Gstix}{\Npar^2 \omegabar\betae} \\ 
\Gstix &= \frac{\sqrt{\pi}\omegape^2 e^{-y^2}}{\abs{\omegace}\kpar\vtherme} 
= \sqrt{\pi}\frac{c^2}{\va^2}\frac{y}{\omegabar}e^{-y^2} \text{ where } y = \frac{\omega}{\kpar\vtherme} 
\end{align}

Note that $\Kyy\antiherm$ will be responsible for transit time magnetic pumping, $\Kzz\antiherm$ for Landau damping, and $\Kyz\antiherm$ for the cross term interaction. Substitution into \eqref{eq:pwave-raw} yields 

\begin{align}
\Pwave &= -\frac{\omega\abs{E_y}^2}{4\sqrt{\pi}}\frac{c^2}{\va^2\omegabar}y e^{-y^2}
\left(\frac{\betae\omegabar\Nperp^2}{2} + \frac{\abs{J}^2}{\betae\omegabar\Npar^2} + \abs{\frac{\kperp}{\kpar}}\Im{J}\right)
\label{eq:Pexp}
\end{align}

We also need the wave energy density $\Wwave$ since $\gammadamp = -\Pwave/\Wwave$. It is defined as 

\begin{align}
\Wwave &= \frac{1}{16\pi}\left[\abs{B}^2 + \vE\conj\dot\pderiv{\left(\omega\Ksym\herm\right)}{\omega}\dot\vE\right]
\end{align}

Above, $\Ksym\herm$ is the Hermitian part of the cold dielectric tensor written in \eqref{eq:wavetens-cold}. For the magnetic field part, use Faraday's Law $c\curl\vE = -\partial\vB/\partial t$. Then this may be evaluated as 

\begin{multline}
\Wwave = \frac{\abs{E_y}^2}{16\pi}\frac{c^2}{\va^2}\left\{N^2(1 + H^2 F^2) 
\vphantom{\frac{A^2}{2}\left[(1+H)^2(1-\omegabar)^2 + (1-H)^2(1+\omegabar)^2\right]}\right. \\ 
+ \left.\frac{A^2}{2}\left[(1+H)^2(1-\omegabar)^2 + (1-H)^2(1+\omegabar)^2\right]\right\}
\label{eq:Wexp}
\end{multline}

Combination of \eqref{eq:Pexp} and \eqref{eq:Wexp}, along with the definitions of $N$ in \eqref{eq:stixdisp}, $H$ in \eqref{eq:Hexp}, and $J$ in \eqref{eq:Jexp}, gives the total damping rate below

\begin{multline}
\frac{\gammadamp}{\omega} = -\frac{4\sqrt{\pi}}{\omegabar}y e^{-y^2} \times \\ 
\frac{\betae\omegabar\Nperp^2/2 + \abs{J}^2/(\betae\omegabar\Npar^2) + \Im{J}\abs{\kperp/\kpar}}
{N^2(1 + H^2 F^2) + A^2\left[(1+H)^2(1-\omegabar)^2 + (1-H)^2(1+\omegabar)^2\right]/2}
\label{eq:dampfull}
\end{multline}

This is a general expression for the total electron damping rate for compressional and shear \Alfven waves when $y \ll 1$ and $\omega \ll \abs{\omegace},\omegape$. It depends on the mode type (compressional vs shear dispersion), frequency $(\omegabar = \omeganorm)$, wave vector direction $(\krat)$, and normalized electron pressure $(\betae)$. In \eqref{eq:dampfull}, the first term in the numerator corresponds to transit time magnetic pumping, the second to Landau damping, and the third to the cross term interaction. 

In order to recover the standard fast wave Landau damping rate in the limit of $\kpar\ll\kperp$, approximate $\Npar^2 \ll \Nperp^2 \approx 1$. Then it follows that $H \approx \omegabar$ and also $J \approx -2i\Npar\Nperp\omegabar\betae$ such that

\begin{align}
\lim_{\kpar\ll\kperp} \frac{\gammadamp^{\text{CAE}}}{\omega} = -\frac{\betae\sqrt{\pi} y e^{-y^2}}{2}
\end{align}

This is the familiar formula from \citeref{Stix1975NF}. The damping rate can also be simplified in the complementary limit of $\kpar \gg \kperp$. In this limit, one can approximate $\Npar^2 \approx 1/(1 \pm \omegabar)$, where the ``$+$'' solution corresponds to CAEs and the ``$-$'' solution is for GAEs. Consequently, $H \approx \pm 1$ and $J \approx -i\omegabar\betae/(2\krat(1 \pm \omegabar)^2)$. Then we find

\begin{align}
\lim_{\kpar\gg\kperp} \frac{\gammadamp}{\omega} = -\frac{\betae\sqrt{\pi} y e^{-y^2}}{2}\frac{\kperp^2}{\kpar^2}\frac{1 \pm 2\omegabar + 2\omegabar^2}{(2\pm\omegabar)(1\pm\omegabar)}
\end{align}

\bibliography{all_bib} 

\end{document}